# Dynamic Self-Consistent Field Theory for Unentangled Homopolymer Fluids

Maja Mihajlovic

*Department of Chemistry, City College, City University of New York, New York, NY 10031, USA*

Tak Shing Lo

*The Levich Institute, City College, City University of New York, New York, NY 10031, USA*

Yitzhak Shnidman\*

*Department of Engineering Science and Physics, College of Staten Island, City University of New York, Staten Island, NY 10314, USA*

November 10, 2004

---

[*]Corresponding author. Electronic address: shnidman@mail.csi.cuny.edu

**ABSTRACT**

We present a lattice formulation of a dynamic self-consistent field (DSCF) theory that is capable of resolving interfacial structure, dynamics and rheology in *inhomogeneous*, *compressible* melts and blends of unentangled homopolymer chains. The joint probability distribution of all the Kuhn segments in the fluid, interacting with adjacent segments and walls, is approximated by a product of one-body probabilities for free segments interacting solely with an external potential field that is determined self-consistently. The effect of flow on ideal chain conformations is modeled with FENE-P dumbbells, and related to stepping probabilities in a random walk. Free segment and stepping probabilities generate statistical weights for chain conformations *in a self-consistent field*, and determine local volume fractions of *chain* segments. Flux balance across unit lattice cells yields mean-field transport equations for the evolution of free segment probabilities and of momentum densities on the Kuhn length scale. Diffusive and viscous contributions to the fluxes arise from segmental hops modeled as a Markov process, with transition rates reflecting changes in segmental interaction, kinetic energy, and entropic contributions to the free energy under flow.

# I. INTRODUCTION

In contrast to equilibrium theories, *nonequilibrium, dynamic* modeling of inhomogeneous polymer fluids is still in its infancy. In such systems, the time evolution of interfacial structure, flow, and rheology is coupled to chain stretching and orientation caused by deformation under flow that is balanced by entropic restoring forces, as well as by segmental interactions. Understanding and predicting the interplay between these processes, and their dependence on composition, and on chain conformation statistics under nonuniform flow, is a challenging fundamental problem. Development of efficient computational models addressing this problem is needed for understanding and design of polymer processing, and for many other industrial and biological applications. We present here a dynamic self-consistent field (DSCF) theory that provides a mean-field approximation for the dynamics of inhomogeneous, unentangled homopolymer fluids, and is capable of resolving coupled interfacial structure, dynamics, and rheology in narrow interfacial regions between macro-phase-separated domains, on the Kuhn length scale.

On small length and time scales, liquids consist of closely packed arrangements of microscopic constituents with a fluctuating distribution of free volume. In simple liquids, the microscopic constituents are atoms or compact molecules [1-4]. The simplest polymer liquids can be viewed as freely jointed chains of Kuhn segments [5-9] that are packed into closely packed arrangements with a fluctuating distribution of free volume. We assume that at any given time, each Kuhn segment is within a cage consisting of closed-packed adjacent segments, and the free volume is distributed as vacancy defects in the close-packed structure. On short time and length scale, we model such cages as Wigner-Seitz unit cells on a close-packed face-centered cubic (FCC) lattice, which are either occupied by a Kuhn segment, or are vacant. Our DSCF theory attributes diffusive and viscous fluxes to segmental hops from occupied cages to adjacent vacant cages. Since it is a formidable task to formulate a continuous model of such hopping events between discrete cages, we chose to use an approximate lattice description. Segmental hopping between adjacent unit cells on a lattice is modeled in our DSCF theory as a Markov stochastic process in time, with thermally activated transition rates [1-4] reflecting changes in segmental interaction, kinetic energy, and entropic chain deformation [6,10] contributions to the free energy under flow, resulting from a single hop. By combining a kinetic mean-field [11] description of the diffusive and viscous contributions to species and momentum fluxes [3,4] with convective and elastic contributions, we obtain microscopic mean-field transport equations for the time evolution of free segment probabilities and momentum densities.

A different Markov process, not to be confused with the one just mentioned above, is used as a tool for characterizing the statistics of chain conformations of jointed chains of Kuhn segments on a lattice, at any fixed moment in time [12]. In this second Markov process, all possible chain conformations are generated by a random walker stepping between adjacent lattice sites, where the steps are labeled not by time, but rather by their order along the generated chain (the so-called "contour length" variable). We reserve the term "hopping" to describe real movement of segments form one cage to another between successive moments in time, as described by the first Markov process. We use the term "stepping" to describe the placement of the next segment at an adjacent lattice site by the fictitious random walker defining the second Markov process.

Consider first a random walker that: (a) has no memory of previously visited sites, (b) is isolated from any other random walkers, walls and external potentials, and (c) does not feel any stretching or orientation constraints imposed by flow. Such a random walker has an isotropic (conditional) stepping probability to step from a site to any adjacent site on the lattice, and generates a distribution of random walks reflecting the statistics of ideal (noninteracting) jointed chains of Kuhn segments in equilibrated polymer melts [5,7]. In the equilibrium SCF lattice theory formulated by Scheutjens and Fleer [12], the interactions between the random walker at a given site, and segments and walls occupying the adjacent sites, are modeled by an effective external potential at the site occupied by the walker. In the presence of this external potential field, which is determined self consistently, the transition rate to make the step toward an adjacent site is modified and becomes the product of the isotropic stepping probability in the absence any potential field, and of the one-body mean-field probability to find a free segment (monomer) interacting with the external potential at the adjacent site.

The mean-field approximation, factorizing joint probabilities for all the Kuhn segments on the lattice into a product of one body probabilities interacting with an external potential field, neglects intrachain correlations between segments jointed into chains. To account for intrachain correlations, the external potential field is calculated self-consistently by averaging over the interactions with walls and with other segments at neighboring sites, using not the one-body free segment probabilities at those sites, but rather the local segmental volume fractions of *connected* segments. The latter are calculated using the master equation governing the random walk in a self-consistent field. Near walls, and in interfacial regions between phase-separated domains of immiscible polymer blends, the segmental volume fractions of *connected* segments are distinct from the one-body free segment probabilities.

If a polymer melt or blend is equilibrated in a channel between two planar walls, which are then subjected to a steady shear, momentum is transferred between the moving walls and the polymer liquid, establishing a nonuniform velocity distribution across the channel which may vary in time. The latter stretches and orients the chains along preferred directions, and thus modifies the statistical distribution of their end-to-end distance at any fixed moment in time. This imposes a constraint on the distribution of chain conformations generated by the random walker. The second moment of the distribution of end-to-end distances of a chain of freely-jointed Kuhn segments is a symmetric second-order tensor called the chain conformation tensor [6,8,9]. In bulk rheology of unentangled polymer fluids, the simplest models for the time evolution of the chain conformation tensor in a nonuniform flow relate it to the time evolution of the second moment of the end-to-end distance of a fictitious dumbbell consisting of just two beads in a nonuniform flow, that are connected by either a harmonic, or a finitely extensible, nonlinearly elastic (FENE) spring [6,8,9], with an entropic (temperature-dependent) spring constant. The latter is tuned to reproduce the equilibrium components of the chain conformation tensor. By matching the components of the chain conformation tensor that is generated by the lattice random walk for ideal chains under flow, with the components of the chain conformation tensor obtained from the FENE model, we are able to determine the values of the anisotropic stepping probabilities that are consistent with the stretching and orientation constraints imposed by the flow.

In continuous versions of equilibrium polymer SCF theory [13-16], the Chapman-Kolmogorov integro-differential equation, rather than the discrete master equation, is the equation governing the random walk Markov process, and chain conformations are generated as continuous trajectories, rather than a sequence of discrete lattice steps. This leads to a time-dependent diffusion equation for the propagation of the random walker in a self-consistent field, where the contour length variable plays the role of time. We chose to use a lattice, rather than a continuum formulation for our polymer DSCF theory because of the inherently discrete nature of the hopping dynamics between close-packed cages with fluctuating free volume, which is difficult to implement in a continuous model. Since a lattice formulation avoids functional differentiations and integrations necessary in a field theoretical continuous formulation, it is also easier to grasp and to simulate.

The evolution equations of our DSCF lattice theory assume the form of a coupled system of nonlinear ordinary differential equations (ODEs), rather than a coupled system of nonlinear partial differential equations (PDEs), which would be the outcome of a continuous theory. This simplifies considerably the analysis, programming, and computational costs involved in numerical solutions of DSCF equations. It also facilitates comparison with molecular dynamics (MD) simulations, which are typically analyzed using averages of densities and fluxes within discrete space-time bins. A continuous, field-theoretical formulation of our DSCF approach is conceivable, but is left for future work.

We conclude this section with an outline for the remainder of the paper. To attain the objectives stated above, the polymer DSCF lattice theory presented here combines many different modeling and approximation methods in an innovative way. In Section II we discuss the physical motivation for each ingredient of our new DSCF theory, and relate it to existing methodologies, before presenting a detailed mathematical formulation. The various ingredients are then linked in a system of coupled nonlinear ODEs evolving a set of independent dynamic variables at each lattice site. The evolution of all other quantities of interest is expressed as functions of these independent dynamic variables. Conclusions and perspectives for application, validation and extension of our DSCF theory are presented in Section III.

## II. MODEL AND FORMULATION

### A. Self-Consistent Mean-Field Approximation

Consider a fluid composed of two molecular species A and B, each species being a linear homopolymer chain of $N^A$ or $N^B$ freely jointed Kuhn segments, respectively. Let the polymer fluid be confined between two parallel solid walls, that are normal to the $z$ axis in an $(x, y, z)$ Cartesian system of coordinates, and are sheared at constant, but opposite velocities $(-u_w, 0, 0)$ and $(u_w, 0, 0)$, as shown in Fig. 1. Such systems have been studied by molecular dynamics (MD) simulations, using both realistic atomic potentials [17], and coarse-grained bead-spring potentials [18-20]. In order to resolve interfacial phenomena, the results of MD simulations are typically analyzed by performing local space and time averages over the MD trajectories ("binning"). The outcome is an evolving set of mean local occupancies and velocities, averaged over a grid of spatial "bins." To resolve steep gradients at interfaces, the spatial dimension of these bins must be comparable to molecular dimensions in the case of simple fluids, and to Kuhn segment dimensions in the case of polymer fluids. The time intervals used for averaging have to be much larger than a single MD time step, but much smaller than the total simulation time. The bin-averaged microscopic evolution arising from MD simulations bears a remarkable resemblance to Eyring's transition-state [1] and Frenkel's kinetic [2] theories of transport in liquids. The bins used in analysis of MD simulations can be associated with molecular or segmental cages moving at a local mean velocity. Small molecules or polymer segments occupying such cages are advected at their local mean velocity, as well as hopping from an occupied cage to an adjacent vacant cage with thermally activated transition rates. Cages may be approximated by lattice unit cells, and the latter identified with the spatial bins used in the binning analysis of MD simulations.

A recent dynamic self-consistent mean-field theory for simple fluids [3,4] models the time evolution of the local mean occupancies and velocities on such a lattice of binning grids. It describes the mass and momentum transport on the cage scale, with mass and momentum fluxes including diffusive and viscous contributions modeled as self-consistent stochastic processes with thermally activated hopping rates, as well as convective contributions. It assumes a mean-field factorization of joint configurational probabilities into a product of one-body probabilities for molecules interacting with a self-consistent field representing adjacent molecules and walls. The polymer DSCF theory presented here aims at a similar objective for unentangled polymer fluids, a task that is complicated by the need to account for intrachain correlations and for deformation of chain conformations under flow.

Motivated by the discussion above, we start the formulation of our DSCF theory by discretizing the space between a pair of parallel sheared walls into unit cells centered about the sites **r** of a face-centered cubic (FCC) lattice [21,22], and forming $L$ triangular lattice layers stacked parallel to the walls, as shown in Fig. 1. Lattice-gas models are widely used to model equilibrium states of simple fluids and are related to the Ising model of ferromagnetism [23]. In lattice-gas models of simple fluids, each unit cell may be occupied by a single molecule. However, flexible linear polymers are approximated by freely-jointed chains of Kuhn segments [7]. The simplest lattice-gas model for polymer fluids, the Flory-Huggins (FH) theory [24-26], assumes that each unit cell is occupied by a Kuhn segment.

Let $a$ be the lattice constant of the FCC lattice. We identify $a$ with the larger of the Kuhn lengths of the two homopolymer species. The volume of the Wigner-Seitz unit cell on the FCC lattice is $w = a^3 / \sqrt{2}$. At any time $t$, each cell represents a segmental cage moving at an instantaneous mean velocity $\mathbf{u_r}$, and is either occupied by a segment of species $\alpha$ (where $\alpha = A, B$), moving at an instantaneous velocity $\mathbf{v_r}$ (which fluctuates about the mean cage velocity $\mathbf{u_r}$), or is vacant. Hence we assign a discrete variable $s_\mathbf{r} = 1, -1$ or $0$ to each lattice site $\mathbf{r}$ to denote whether it is occupied by a segment of species $A$, a segment of species $B$, or vacant, respectively. Note that the presence of vacancies makes our DSCF model for polymer fluids *compressible*. In this respect it is different from the original Scheutjens and Fleer's lattice SCF theory for equilibrium polymer fluids [27], but is similar to its compressible variants [28-30].

Let $\mathbf{e}_1 = (1, 0, 0)$, $\mathbf{e}_2 = (0, 1, 0)$ and $\mathbf{e}_3 = (0, 0, 1)$ be a triad of unit vectors serving as the basis for the $(x, y, z)$ Cartesian system of coordinates. A site $\mathbf{r} = (x, y, z)$ belonging to the FCC lattice has 12 nearest-neighbors at $\mathbf{r} + \mathbf{a}_k$, where

$$\begin{aligned}
\mathbf{a}_1 &= a(1, 0, 0) = -\mathbf{a}_4, & \mathbf{a}_2 &= a\left(\tfrac{1}{2}, \tfrac{\sqrt{3}}{2}, 0\right) = -\mathbf{a}_5, & \mathbf{a}_3 &= a\left(-\tfrac{1}{2}, \tfrac{\sqrt{3}}{2}, 0\right) = -\mathbf{a}_6, \\
\mathbf{a}_7 &= a\left(0, -\tfrac{1}{\sqrt{3}}, \sqrt{\tfrac{2}{3}}\right) = -\mathbf{a}_{10}, & \mathbf{a}_8 &= a\left(\tfrac{1}{2}, \tfrac{1}{2\sqrt{3}}, \sqrt{\tfrac{2}{3}}\right) = -\mathbf{a}_{11}, & \mathbf{a}_9 &= a\left(-\tfrac{1}{2}, \tfrac{1}{2\sqrt{3}}, \sqrt{\tfrac{2}{3}}\right) = -\mathbf{a}_{12}.
\end{aligned} \tag{1}$$

The six sites $\{ \mathbf{r} + \mathbf{a}_1, \mathbf{r} + \mathbf{a}_2, \mathbf{r} + \mathbf{a}_3, \mathbf{r} + \mathbf{a}_4, \mathbf{r} + \mathbf{a}_5, \mathbf{r} + \mathbf{a}_6 \}$ form a nearest-neighbor shell on a triangular lattice within the $(x, y)$ plane at distance $z$ from the origin. The three lattice sites $\{ \mathbf{r} + \mathbf{a}_7, \mathbf{r} + \mathbf{a}_8, \mathbf{r} + \mathbf{a}_9 \}$, stacked as shown in Fig. 2, occupy another triangular lattice on at the $(x, y)$ plane at a distance $z + \sqrt{\frac{2}{3}} a$ from the origin. Similarly, the three lattice sites $\{ \mathbf{r} + \mathbf{a}_{10}, \mathbf{r} + \mathbf{a}_{11}, \mathbf{r} + \mathbf{a}_{12} \}$, stacked as shown in Fig. 2, form another triangular lattice on the plane $(x, y)$ at a distance $z - \sqrt{\frac{2}{3}} a$ from the origin. Our objective is to obtain the time evolution of all relevant dynamic variables defined at the sites of the FCC lattice belonging to $L$ triangular lattice layers parallel to the walls, at $z_i = \sqrt{\frac{2}{3}} i a$ (where $i = 1, \ldots, L$). The bottom wall is located at the plane $z = 0$, and the sites of the triangular lattice layer adjacent to the bottom wall are located in the plane $z = z_1 \equiv \sqrt{2/3} a$. Similarly, the top wall is at the plane $z = \sqrt{2/3} a (L+1)$, and the sites of the triangular lattice layer adjacent to it are in the plane $z = z_L \equiv \sqrt{2/3} a L$. We assume periodic boundary conditions within the triangular lattice layers.

We assume that at any given time $t$, any lattice site $\mathbf{r}$ between the two walls is: (a) occupied by a Kuhn segment of type $A$, in which case it is assigned an occupancy (pseudospin) variable $\sigma_{\mathbf{r}} = 1$, or (b) occupied by a Kuhn segment of type $B$, in which case it is assigned an occupancy (pseudospin) variable $\sigma_{\mathbf{r}} = -1$, or (c) vacant, in which case it is assigned an occupancy (pseudospin) variable $\sigma_{\mathbf{r}} = 0$. Segments or vacancies occupying this site have an instantaneous velocity $\mathbf{v}_{\mathbf{r}}$. The microscopic state of the system at time $t$ is specified by the values of the occupancy (pseudospin) variables $\sigma_{\mathbf{r}}$ and the segmental velocities $\mathbf{v_r}$ at each lattice site $\mathbf{r}$ between the two walls. Denoting the collection of all the sites between the two walls by $\Omega$, the microscopic state is assigned an (exact) phase-space probability $P_{ex}\left(\{\sigma_{\mathbf{r}}, \mathbf{v_r}\}_{\mathbf{r}\in\Omega}, t\right)\prod_{\mathbf{r}\in\Omega} d\mathbf{v_r}$.

Self-consistent field (SCF) theories *approximate* the configuration probability for a many-body system with a *product of one-body probabilities* for each constituent, interacting solely with a mean external potential field. This field represents a configurational energy of interaction with all the other system constituents. In the case of lattice-gas models of simple fluids, and other Ising-like systems, this interaction energy is averaged over the *one-body probability distributions* of the remaining constituents. For lattice-gas models of a compressible liquid at equilibrium the simplest SCF theory is equivalent to the Bragg-Williams theory for the Ising ferromagnet [23,31-33]. An early example of a dynamic mean-field theory is Boltzmann's kinetic equation [34] for the time evolution of one-body probability distributions in gases, where products of one-body probability approximate pair probabilities. Lattice-Boltzmann methods [35-39] are lattice versions of Boltzmann's kinetic equation. In the continuum limit, they reproduce the description of transport in inhomogeneous fluids by means of partial differential equations (such as the Navier-Stokes, Cahn-Hilliard-Cook [40,41], and time-dependent Landau Ginzburg [21,42] equations), but are computationally more efficient. The same assumption about factorization of correlations plays a key role in the derivation of kinetic mean-field equations from the underlying master equation governing the time evolution of configurational probability in stochastic lattice-gas models [11].

Mean-field theories for polymer fluids treat Kuhn segments as system constituents. They *approximate* the configuration probability for a many-segment system with a *product of one-body probabilities* for each segment, interacting solely with a conjugate mean external potential field, which is the prescription that we follow here:

$$P_{ex}\left(\left\{\sigma_{\mathbf{r}}, \mathbf{v}_{\mathbf{r}}\right\}_{\mathbf{r}\in\Omega}, t\right)\prod_{\mathbf{r}\in\Omega} d\mathbf{v}_{\mathbf{r}} = \prod_{\mathbf{r}\in\Omega} P\left(\sigma_{\mathbf{r}}, t\right)Q\left(\sigma_{\mathbf{r}}, \mathbf{v}_{\mathbf{r}}, t\right)d\mathbf{v}_{\mathbf{r}} \qquad (2)$$

Here $P\left(\sigma_{\mathbf{r}}, t\right)$ are time dependent one-body probabilities for free segments (monomers) of type $A$, or $B$, or vacancies, to be at lattice sites $\mathbf{r}$, where they interact solely with a local mean potential field. The latter represent interactions of a segment at a given site with segments or walls at adjacent sites. Note that our model is *compressible*, since we allow vacancies at lattice sites. Similarly $Q\left(\sigma_{\mathbf{r}}, \mathbf{v}_{\mathbf{r}}, t\right)d\mathbf{v}_{\mathbf{r}}$ is the one-body probability for the velocities $\mathbf{v}_{\mathbf{r}}$ of the respective free segments or vacancies at lattice sites $\mathbf{r}$ to be in a volume element $d\mathbf{v}_{\mathbf{r}}$. A similar factorization approximation was used previously in a DSCF mean field study of sheared *simple* fluids [3,4]. However, in that study, each cage was occupied by a compact atom or molecule, rather than by a Kuhn segment of a polymer chain.

Note that Eq. (2) neglects any correlations between clusters of free segments. Thus, in general, $P\left(\sigma_{\mathbf{r}}, t\right)$ is different from the local segmental volume fraction $\phi_{\mathbf{r}}^{\alpha}\left(t\right)$ of a polymer species at site $\mathbf{r}$, which reflects the effect of intrachain correlations between segments arising from chain connectivity. Eq. (2) also neglects any correlations arising from interactions of a segment at a given site with segments or walls at any other site. These are modeled as interactions with a local, time-dependent consistent field at the given site. We refer to $P\left(\sigma_{\mathbf{r}}, t\right)$ as the time-dependent *free segment probability* at site $\mathbf{r}$, since under equilibrium conditions it reduces to the time-independent free segment (monomer) probability originally proposed by Scheutjens and Fleer in their lattice SCF theory for inhomogeneous polymer fluids at thermodynamic equilibrium [27].

Henceforth we use a local equilibrium approximation for the velocity probability distribution function $Q(\sigma_{\mathbf{r}}, \mathbf{v}_{\mathbf{r}}, t)$. This means that locally it has the same form as the equilibrium Maxwell-Boltzmann distribution of segmental velocities, in the frame of reference of a cage centered at site $\mathbf{r}$ and moving at a mean velocity $\mathbf{u}_{\mathbf{r}}$, and that vacancies are advected at the mean cage velocity $\mathbf{u}_{\mathbf{r}}$, as follows,

$$Q(\sigma_{\mathbf{r}}, \mathbf{v}_{\mathbf{r}}, t) = \begin{cases} \left(2\pi m_s^A k_B T\right)^{-3/2} \exp\left[-m_s^A \left(\mathbf{v}_{\mathbf{r}} - \mathbf{u}_{\mathbf{r}}\right)^2 / 2k_B T_{\mathbf{r}}\right] & \text{if} \quad \sigma_{\mathbf{r}} = 1, \\ \left(2\pi m_s^B k_B T\right)^{-3/2} \exp\left[-m_s^B \left(\mathbf{v}_{\mathbf{r}} - \mathbf{u}_{\mathbf{r}}\right)^2 / 2k_B T_{\mathbf{r}}\right] & \text{if} \quad \sigma_{\mathbf{r}} = -1, \\ \delta\left(\mathbf{v}_{\mathbf{r}} - \mathbf{u}_{\mathbf{r}}\right) & \text{if} \quad \sigma_{\mathbf{r}} = 0. \end{cases} \tag{3}$$

The local equilibrium approximation in Eq. (3) neglects the dependence of the velocity probability distribution function on other (second and higher) moments of the velocity. Accounting for this dependence requires better approximations, such as Grad's "thirteen moments" method [10,34]. For the remainder of this paper, we assume that the system is isothermal, so that at all times $t$, $T_{\mathbf{r}} = T$ for all $\mathbf{r} \in \Omega$, though in the case of simple fluids it has been shown how this assumption can be relaxed to model nonisothermal and Marangoni flows [4]. Since $Q(\sigma_{\mathbf{r}}, \mathbf{v}_{\mathbf{r}}, t)$ are simple Gaussian and $\delta$ functions of $\mathbf{v}_{\mathbf{r}}$, velocity moments such as $\int_{-\infty,-\infty,-\infty}^{\infty,\infty,\infty} \mathbf{v}_{\mathbf{r}}^n Q(\sigma_{\mathbf{r}}, \mathbf{v}_{\mathbf{r}}, t) d\mathbf{v}_{\mathbf{r}}$ are easily evaluated functions of the mean cage velocities $\mathbf{u}_{\mathbf{r}}$ and the temperature $T$ that no longer depend on $\mathbf{v}_{\mathbf{r}}$.

To simplify the form of the equations, we henceforth drop the explicit time dependence of various quantities, and denote $P(\sigma_{\mathbf{r}}, t) = P_{\mathbf{r}}^A$ or $P_{\mathbf{r}}^B$ when site $\mathbf{r}$ in layer $i$ is occupied by a segment of species A or B ($\sigma_{\mathbf{r}} = \pm 1$), respectively, and $P(\sigma_{\mathbf{r}}, t) = 1 - P_{\mathbf{r}}^{A} - P_{\mathbf{r}}^{B}$ when this site is vacant ($\sigma_{\mathbf{r}} = 0$). The free segment probabilities $P_{\mathbf{r}}^{\alpha}$ relate to $\tilde{P}_{\mathbf{r}}^{\alpha}$, the statistical weights to place a free segment of species $\alpha$ at a site in layer $i$, as

$$P_{\mathbf{r}}^{\alpha} = \tilde{P}_{\mathbf{r}}^{\alpha} / \left(1 + \tilde{P}_{\mathbf{r}}^{A} + \tilde{P}_{\mathbf{r}}^{B}\right), \tag{4}$$

and hence

$$\tilde{P}_{\mathbf{r}}^{\alpha} = P_{\mathbf{r}}^{\alpha} / \left(1 - P_{\mathbf{r}}^{A} - P_{\mathbf{r}}^{B}\right). \tag{5}$$

According to Eq. (5) a vacancy at a site is always assigned a constant statistical weight of 1 (since vacancies cannot interact with a self-consistent field), while the statistical weights for a free segment of type A or B to occupy a site can be a non-negative number larger or smaller than 1 (reflecting either an attractive or a repulsive self-consistent potential for free segments of this type occupying the site $\mathbf{r}$, respectively). Eq. (4) just states that the free segment probability for a segment of type A or B to occupy site $\mathbf{r}$ equals the statistical weight for this event, normalized by the sum of the statistical weights for all three possible outcomes (segment of type A, segment of type B, or vacancy) at this site. Hence the probabilities for all three possible outcomes at the site sum up to 1, as they should. This assignment of statistical weights for free segments is consistent with the one used by Scheutjens and Fleer in their original derivation of *equilibrium* SCF lattice theory for *incompressible* polymer fluids [12].

### B. Random Walk Model for Chain Conformations and Intrachain Correlations

Note that Eq. (2) neglects any correlations between clusters of free segments, and, in particular, the intrachain correlations arising from chain connectivity. In FH theory [24-26], the translational entropy per segment in a system containing chains of $N^{\alpha}$ *connected* segments is reduced by a factor of $1/N^\alpha$ compared to a system containing free segments (monomers). This is because segments connected in a chain are correlated: if one segment in a chain is translated, all the other segments belonging to the same chain are translated along with it. However, this is not necessarily true if the chain changes its conformation as it is being translated, an effect that is neglected in the FH approximation. FH theory was used extensively to construct equations of state and phase diagrams of *homogeneous* polymer fluids.

In general, unfavorable interactions with walls, or with segments belonging to a different species at an interface between two coexisting phases, may lead to further corrections to the mean-field chain entropy in interfacial regions. Accounting for such corrections requires a more accurate treatment of the intrachain correlations and of the statistics of chain conformations than in FH theory [43]. This can be achieved if, in addition to the many-body interactions, the intrachain correlations between the Kuhn segments and the deviations of chain conformations from ideality in the interfacial regions are treated self-consistently as well. In line with the prevalent terminology of polymer physics, the term "polymer SCF theory" is reserved for such "doubly self-consistent" methods, to distinguish them from FH or related theories, in which only configurational interactions between segments are treated self-consistently.

The starting point of SCF theories for inhomogeneous polymer fluids at equilibrium is the one-body probability $P_\mathbf{r}^\alpha$ for a free Kuhn segment (monomer) in a self-consistent mean potential field. The latter represents interactions with adjacent segments belonging to connected chains and with walls. In the lattice SCF theory of Scheutjens and Fleer [27] it is called the free segment probability. In the continuous version of polymer SCF theory [13-16] it provides the initial condition for a quasi-time-dependent diffusion equation for the local probability of chain termini, where the chain contour length plays the role of time. As the free segment probability neglects intrachain correlations between segments, its value in interfacial regions is distinct from the local segmental volume fraction $\phi_{\mathbf{r}}^{\alpha}$ of *connected* segments.

The dependence of $P_{\mathbf{r}}^{\alpha}$ on the local, self-consistent, mean potential field reflects interactions with segments or walls at adjacent sites. Under nonequilibrium conditions, both $P_{\mathbf{r}}^{\alpha}$ and the self-consistent field are in general time-dependent, and the self-consistent field should also reflect chain stretching and orientation effects caused by nonuniform flows. To account for intrachain correlations, the self-consistent field at site $\mathbf{r}$ must depend on the local segmental volume fractions $\phi_{\mathbf{r}}^{\alpha}$ and $\phi_{\mathbf{r}+\mathbf{a}_k}^{\alpha}$ for segments of type $\alpha$ belonging to a *connected* chain of $N^{\alpha}$ freely jointed segments of type $\alpha$ to be at site $\mathbf{r}$ and the adjacent sites $\mathbf{r}+\mathbf{a}_k$, where $k = 1,...,12$. If we assumed, instead, that it depends on the free segment probabilities $P_{\mathbf{r}}^{\alpha}$ and $P_{\mathbf{r}+\mathbf{a}_k}^{\alpha}$, we would have neglected intrachain correlations entirely, resulting in a much worse approximation for inhomogeneous polymer fluids. Following Scheutjens and Fleer [27], we account for such correlations by representing chain conformations as lattice random walks in a self consistent field.

In an unentangled, homogeneous melt phase of identical homopolymer chains at thermodynamic equilibrium, excluded volume interactions between segments can be neglected. This is because excluded volume interactions between segments belonging to the same chain are screened out by the excluded volume interactions with segments belonging to the other chains (Flory's theorem [7]). Thus such ideal (noninteracting) chains are modeled as isotropic lattice random walks [44,45] placing noninteracting segments at successive lattice sites, with the chain contour length variable $s$ playing the role of time [27]. All possible chain conformations of an ideal chain of $N^\alpha$ segments, starting with the first segment being at a given site, are generated recursively with equal statistical weight. At equilibrium, this is done by selecting any of the sites adjacent to the terminus of a chain of $s$ segments with isotropic (equal) single-step displacement ("stepping") probability [44,45] $\lambda = 1/q$, where $q$ is the lattice coordination number ($q = 12$ for the FCC lattice). A segment is then placed at this adjacent site and connected to the terminus of the $s$-segment chain, producing a conformation of a chain of $s+1$ connected segments. The process is repeated recursively, until a conformation of a chain of $N^\alpha$ jointed Kuhn segments is produced. Such a random walk is a Markov chain obeying a master equation [44,45]. The continuum limit of the master equation has the form of a time dependent diffusion equation with an isotropic diffusion coefficient, where $s$ plays the role of time. The solution of the time-dependent diffusion equation subject to a delta-function initial condition has the form of an isotropic Gaussian with a second moment of the end-to-end distance that is proportional to $N^\alpha - 1$ [45].

In equilibrium SCF theory for polymer fluids, there are several contributions to the local self-consistent mean potential field for the one-body free segment probability. One contribution is from the configurational interactions of the segment with adjacent segments, solvent molecules, and walls, averaged over $\phi_r^\alpha$, the local *segmental volume fractions* at adjacent sites, rather than over $P_r^\alpha$, the one-body free segment probabilities. Another contribution originates from effect of intrachain correlations on configurational entropy. If it is assumed that the polymer fluid is incompressible, as in most applications of polymer SCF theory to date, there is another contribution to the self consistent field that imposes the incompressibility constraint. A compressible version of lattice SCF theory for polymer fluids at equilibrium has been formulated by Theodorou [28], combining ideas from Scheutjens and Fleer's lattice SCF theory [27] with the equation of state theory of Sanchez and Lacomb [46,47]. Here we adopt a simpler way to model compressible polymer fluids by introducing a noninteracting monomer solvent species representing vacancies into Scheutjens and Fleer's SCF theory [30].

The new feature of our DSCF theory is that here the two parallel walls are allowed to move at different velocities, driving the fluid out of thermodynamic equilibrium. Thus all quantities of interest may depend on time. Momentum transfer at the walls may induce a time-dependent, nonuniform velocity field $\mathbf{u_r}$, representing the mean cage velocities at sites $\mathbf{r}$ [3,4], which is a lattice discretization of the continuous nonuniform velocity field in the real fluid. Such a nonuniform flow field may stretch and orient the polymer chains along preferred directions. As a result, the second moment of the chain end-to-end distance becomes anisotropic even in a melt of *ideal* chains of homogeneous density [6,10,48]. We account for such anisotropic stretching by allowing $\lambda_{\mathbf{a}_k,\mathbf{r}}^{\alpha}$, the stepping probability (from a segment belonging to an *ideal* chain that is located at site $\mathbf{r}$ to a connected segment at the adjacent site $\mathbf{r}+\mathbf{a}_k$ ), to be anisotropic and dependent on both position and time. Hence we replace the equilibrium assumption that the stepping probabilities $\left\{\lambda_{\mathbf{a}_k,i}^{\alpha}\right\}$ are isotropic (identical for all $k$ ) [27], with a less restrictive reflection symmetry. Explicitly, we assume that at any site $\mathbf{r}$ belonging to the FCC lattice of Fig. 2,

$$\lambda_{\mathbf{a}_l,\mathbf{r}}^{\alpha} = \lambda_{\mathbf{a}_k,\mathbf{r}}^{\alpha} \quad \text{if} \quad \mathbf{a}_l = -\mathbf{a}_k . \tag{6}$$

Thus of the twelve stepping probabilities from a site toward its nearest neighbors on the FCC lattice, only six are independent.

At any given time, the statistical weights for conformations of ideal chains generated by such an anisotropic random walk [45] procedure will depend on the direction of the steps taken to generate this particular conformation. All possible conformations of chains in a flow field, which interact with other chains or with the walls, are similarly generated by a random walk in a self-consistent mean potential field representing such many-body interactions. The statistical weight to find the terminus of a chain of $N^\alpha$ connected Kuhn segments at any site is found recursively following the algorithm specified by Scheutjens and Fleer [27], except the stepping probabilities for steps towards different nearest neighbors on the FCC lattice are now anisotropic (depend on $k$), subject to the reflection symmetry constraint, Eq. (6). Let $P^\alpha_{\mathbf{r}+\mathbf{a}_k}(s)$ be the statistical weight to find a terminus of a chain of $s$ connected segments of type $\alpha$ at any of the sites $\mathbf{r}+\mathbf{a}_k$ that are adjacent to a given site $\mathbf{r}$. The transition rate to connect an additional segment of type $\alpha$ at site $\mathbf{r}$ to the terminus of an $s$-segment chain at an adjacent site $\mathbf{r}+\mathbf{a}_k$ is the product of the statistical weight $\tilde{P}^\alpha_{\mathbf{r}}$ for a free segment of type $\alpha$ to be at site $\mathbf{r}$ and an anisotropic stepping probability $\lambda^\alpha_{-\mathbf{a}_k,\mathbf{r}+\mathbf{a}_k}$. The value of $\lambda^\alpha_{\mathbf{a}_k,\mathbf{r}}$ for each bond in the $\mathbf{a}_k$ direction emanating from site $\mathbf{r}$ is represented by the shading of that bond in Fig. 3, and the value of $\tilde{P}^\alpha_{\mathbf{r}}$ at site $\mathbf{r}$ is represented by the background shading. Fig. 3 is a schematic representation of a chain conformation generated on a triangular, rather than the FCC lattice, for simplicity reasons. Thus the time-dependent statistical weight $P^\alpha_{\mathbf{r}}(s+1)$ for finding a terminus of an arbitrary $\alpha$-type subchain of length $s+1$ at the lattice site $\mathbf{r}$ is defined recursively as follows [49]

$$P_{\mathbf{r}}^{\alpha}\left(s+1\right) = \tilde{P}_{\mathbf{r}}^{\alpha} \sum_{k=1}^{12} \lambda_{-\mathbf{a}_k,\mathbf{r}+\mathbf{a}_k}^{\alpha} P_{\mathbf{r}+\mathbf{a}_k}^{\alpha}\left(s\right) \qquad (7)$$

with the initial condition $P_{\mathbf{r}}^{\alpha}\left(1\right) = \tilde{P}_{\mathbf{r}}^{\alpha}$. Eq. (7) is a master equation for the evolution of terminal probabilities $P_{\mathbf{r}}^{\alpha}\left(s\right)$, where the contour length variable $s$ plays the role of time. In the continuous version of polymer SCF theory [13-15] it is replaced by a Chapman-Kolmogorov integro-differential equation [44].

Note that for an ideal, non-interacting chain of type $\alpha$, $\tilde{P}_{\mathbf{r}}^{\alpha} = 1$. In this case the background is uniform (shown as a white background in Figs. 3(a) and 3(b)). If, in addition, the chain and the surrounding fluid are in thermodynamic equilibrium, the values of all the stepping probabilities used in generating the chain are isotropic (independent of bond directions), as represented by identical bond shading for bonds along different directions in Fig. 3(a). The second moment of the distance between the two ends of the chain (denoted by the zigzag line connecting two solid circles) is isotropic (denoted by the dashed circle in Fig. 3(a) corresponding to a sphere on the FCC lattice). On the other hand, if the fluid has homogeneous composition, but the velocity of the surrounding fluid is nonuniform, stretched and oriented conformations will be assigned higher statistical weights because of higher values of $\lambda_{\mathbf{a}_k,\mathbf{r}}^{\alpha}$ along preferred bond directions. This is shown by different shading of bonds emanating from different sites $\mathbf{r}$ along different directions $\mathbf{a}_k$. In this case the second moment of the chain end-to-end is stretched and anisotropic (denoted by a dashed ellipse in Fig. 3b, representing an ellipsoid on the FCC lattice). If, on top of that, the fluid has an inhomogeneous composition, the values of $\tilde{P}_{\mathbf{r}}^{\alpha}$ at different sites will be different (as shown by the nonuniform shading of the background in Fig. 3(c)).

The volume fractions $\phi_{\mathbf{r}}^{\alpha}$ of segments of type $\alpha$ occupying a site $\mathbf{r}$ at a given time are calculated from known values of $P_{\mathbf{r}}^{\alpha}(s)$ at that time, as follows [27],

$$\phi_{\mathbf{r}}^{\alpha} = C_{\alpha} \sum_{s=1}^{N^{\alpha}} \frac{P_{\mathbf{r}}^{\alpha}(s) P_{\mathbf{r}}^{\alpha}(N^{\alpha} - s + 1)}{\tilde{P}_{\mathbf{r}}^{\alpha}}. \qquad (8)$$

According to Eq. (8), first the statistical weight for having the $s$-segment of an $N^{\alpha}$-segment chain in layer $i$ is calculated. It is expressed as a product of the statistical weights for two sub-chains, one of length $s$, and another of length $(N^{\alpha} - s + 1)$, to terminate at a common site in layer $i$, which is then divided by $\tilde{P}_{\mathbf{r}}^{\alpha}$ to compensate for double counting in the placement of the terminal segment common to the two sub-chains. To get $\phi_{i}^{\alpha}$, these statistical weights are then summed over all possible values of $s$ along the chain, and normalized. In the continuous version of polymer SCF theory [13-15], the sum over discrete values of $s$ is replaced by an integral over a continuous $s$ variable. The normalization constant is

$$C_{\alpha} = \frac{\overline{\phi}^{\alpha} n_{s}}{N^{\alpha} \sum_{\mathbf{r}} P_{\mathbf{r}}^{\alpha}(N^{\alpha})}, \qquad (9)$$

where $n_{s}$ is the total number of sites in the system, and $\overline{\phi}^{\alpha}$ is the mean segmental fraction of species $\alpha$:

$$\overline{\phi}^{\alpha} = \frac{1}{n_{s}} \sum_{\mathbf{r}} \phi_{\mathbf{r}}^{\alpha}. \qquad (10)$$

Note that $\overline{\phi}^\alpha$ is constant in a non-reacting closed system between impermeable walls. For a two-component A/B blend we define $\phi_\mathbf{r}$, the total segmental volume fraction at site $\mathbf{r}$, and $\overline{\phi}_\mathbf{r}$, the total mean segmental volume fraction, as follows,

$$\phi_\mathbf{r} = \phi_\mathbf{r}^A + \phi_\mathbf{r}^B,$$
$$\overline{\phi} = \overline{\phi}^A + \overline{\phi}^B. \qquad (11)$$

Let $m^\alpha$ be the mass of a segment of species $\alpha$. Then the mass density of species $\alpha$ at a site belonging to layer $i$ is $\rho_\mathbf{r}^\alpha = m^\alpha \phi_\mathbf{r}^\alpha w^{-1}$. Let

$$\mathbf{g}_\mathbf{r}^\alpha = \sum_{k=1}^{3} g_{\mathbf{r},k}^\alpha \mathbf{e}_k \qquad (12)$$

be the momentum density of connected segments belonging to chains of type $\alpha$ at that site, then the local mean velocity of species $\alpha$ is given by

$$\mathbf{u}_\mathbf{r}^\alpha = \sum_{k=1}^{3} u_{\mathbf{r},k}^\alpha \mathbf{e}_k \equiv \left( \frac{w}{m^\alpha \phi_\mathbf{r}^\alpha} \right) \mathbf{g}_\mathbf{r}^\alpha = \frac{\mathbf{g}_\mathbf{r}^\alpha}{\rho_\mathbf{r}^\alpha}. \qquad (13)$$

We identify the local mean cage velocity with the mass-averaged mean velocity at a site in the same layer, defined by

$$\mathbf{u}_\mathbf{r} = \frac{\mathbf{g}_\mathbf{r}}{\rho_\mathbf{r}}. \qquad (14)$$

Here $\mathbf{g}_\mathbf{r} = \mathbf{g}_\mathbf{r}^A + \mathbf{g}_\mathbf{r}^B$ is the total momentum density and $\rho_\mathbf{r} = \left( m^A \phi_\mathbf{r}^A + m^B \phi_\mathbf{r}^B \right) / w$ is the total mass density at site $\mathbf{r}$.

## C. Evolution of Anisotropic Stepping Probabilities

At any moment in time, the modified Scheutjens-Fleer procedure outlined above allows determination of the local segmental volume fractions $\phi_\mathbf{r}^\alpha$ for all sites $\mathbf{r}$, provided that we know not only the free segment probabilities $P_\mathbf{r}^\alpha$, but also the anisotropic stepping probabilities $\lambda_{\mathbf{a}_{k},\mathbf{r}}^{\alpha}$ representing chain stretching and orientation by nonuniform flows. The $\lambda_{\mathbf{a}_{k},\mathbf{r}}^{\alpha}$ are the conditional stepping probabilities to place the next segment in a connected chain at any of the 12 sites adjoining the site $\mathbf{r}$, where the current segment is located. They are constant and isotropic in the equilibrium SCF theory of Scheutjens and Fleer [27]. In the DSCF theory presented here, $\lambda_{\mathbf{a}_{k},\mathbf{r}}^{\alpha}$ are anisotropic and time dependent functions, related to stretching and orientation of ideal (noninteracting) chains in a homogeneous fluid induced by nonuniform flows, characterized by a time dependent velocity field $\mathbf{u_r}$ (see Fig. 3(b)).

As stated previously, a homogeneous, one-component homopolymer melt, or a polymer solution near its theta-point, can be considered as a fluid of ideal chains where configurational interactions between segments are negligible. However the statistical weights for ideal chains are perturbed from their equilibrium values when subjected to flow velocity gradients (see Eq. (7) and Fig. 3(b)). Their time evolution should reflect chain stretching and orientation under nonuniform flow, as well as their entropically-driven relaxation toward thermodynamic equilibrium. Bead-spring models of such phenomena were constructed for modeling the rheology of unentangled, homogeneous polymer fluids [5,6,8,9]. Homopolymer chain consisting of $N^{\alpha}$ identical, freely jointed Kuhn segments in an unentangled homogeneous melt, can be represented by the Rouse model [50], consisting of $N^{\alpha}$ beads connected by entropic springs, subject to friction and random forces exerted by the surrounding fluid.

Here we use even simpler elastic dumbbell [6] models for the time evolution of the second moment $\mathbf{S_r^{\alpha}} = \left\langle \mathbf{Q_r^{\alpha} Q_r^{\alpha}} \right\rangle$ of the end-to-end distance $\mathbf{Q_r^{\alpha}}$ in a homogeneous fluid of ideal chains (in Fig. 3, solid circles represent the two beads and the zigzag line represents the elastic spring). A Hookean dumbbell consisting of two beads connected by a linear spring [51,52] has unbounded end-to-end distance, which is not realistic. Here we use a dumbbell model with a finitely extensible, nonlinear spring in the Peterlin approximation (FENE-P) [53-58], where the nonlinear spring force is *pre-averaged* over the probability distribution of end-to-end distances. Thus in our DSCF theory, the effect of flow on conformations of ideal chains is computed using the FENE-P dumbbell model for the time evolution of $\mathbf{S}_{\mathbf{r}}^{\alpha}$. We show below how $\mathbf{S}_{\mathbf{r}}^{\alpha}$ is related to $\lambda_{\mathbf{a}_s,\mathbf{r}}^{\alpha}$. At any given time, the recursive relation (7) uses known values of $P_{\mathbf{r}}^{\alpha}$ and $\lambda_{\mathbf{a}_s,\mathbf{r}}^{\alpha}$ to generate values of $P_{\mathbf{r}}^{\alpha}\left(s\right)$, which are then substituted into Eq. (8) to compute the values of $\phi_{\mathbf{r}}^{\alpha}$. We will show below that $P_{\mathbf{r}}^{\alpha}$ depends on a self-consistent, time-dependent, mean potential field at $\mathbf{r}$, which models not only segment-segment and segment-wall interactions, but also changes in entropy due to chain deformations under flow. Thus our Eq. (7) accounts not only for interactions with adjacent segments and walls [27] in an inhomogeneous fluid, but also for entropy changes due to chain stretching and orientation under nonuniform flows. Chain deformation caused by segment-segment and segment-wall interactions is modeled by the dependence of the statistical weights for chain conformations in Eq. (7) on $P_{\mathbf{r}}^{\alpha}$, as in the original equilibrium SCF model of Scheutjens and Fleer [27].

Let $\mathbf{Q}_{\mathbf{r}}^{\alpha}$ be the end-to-end distance in a chain of $N^{\alpha}$ freely jointed Kuhn segments of length $a$ each. We use a FENE-P dumbbell model for the time evolution of the probability distribution of $\mathbf{Q}_{\mathbf{r}}^{\alpha}$ and its second moment. In this model, $\mathbf{Q}_{\mathbf{r}}^{\alpha}$ is the distance between the two dumbbell beads (representing the two ends of the freely jointed chain of

$N^\alpha$ Kuhn segments), connected by an entropic spring with a spring constant given by

$3k_B T \left[ \left( N^\alpha - 1 \right) \tilde{a}_\alpha^2 \left( 1 - \left\langle \left( \mathbf{Q}_\mathbf{r}^\alpha \right)^2 \big/ Q_0^2 \right\rangle \right) \right]^{-1}$. Here $Q_0$ is the maximum extension of the spring

[53-58], and $\tilde{a}_\alpha = c_\alpha a$ is a length proportional to the Kuhn length. We approximate the

probability to find a FENE-P dumbbell (in an ideal fluid of noninteracting FENE-P

dumbbells), with its center of mass at site $\mathbf{r}$, and the distance $\mathbf{Q}_\mathbf{r}^\alpha$ between the two beads,

by the product $n_\mathbf{r}^\alpha \psi_\mathbf{r}^\alpha \left( \mathbf{Q}_\mathbf{r}^\alpha \right)$, where $n_\mathbf{r}^\alpha = \left( \phi_\mathbf{r}^\alpha w^{-1} \right) \big/ N^\alpha$ is the number of chains composed of segments of type $\alpha$ per volume of a unit cell centered at site $\mathbf{r}$.

The time evolution of the probability distribution $\psi_\mathbf{r}^\alpha \left( \mathbf{Q}_\mathbf{r}^\alpha \right)$ is described by a

Smoluchowski equation [6,58]. The time evolution of $\mathbf{S}_\mathbf{r}^\alpha = \left\langle \mathbf{Q}_\mathbf{r}^\alpha \mathbf{Q}_\mathbf{r}^\alpha \right\rangle$ is derived by

evaluating the second moment of each term in the Smoluchowski equation, resulting in

the following equation [6,58]:

$$\overset{\triangledown}{\mathbf{S}}_\mathbf{r}^\alpha = -\frac{1}{\tau_{db,\mathbf{r}}^\alpha} \left[ \frac{\mathbf{S}_\mathbf{r}^\alpha}{1 - \frac{3}{\left( N^\alpha - 1 \right) \tilde{a}_\alpha^2 b^\alpha} \mathrm{Tr} \mathbf{S}_\mathbf{r}^\alpha} - \frac{\left( N^\alpha - 1 \right) \tilde{a}_\alpha^2}{3} \boldsymbol{\delta} \right], \tag{15}$$

where $\overset{\triangledown}{\mathbf{S}}_\mathbf{r}^\alpha = D\mathbf{S}_\mathbf{r}^\alpha / Dt - \left( \nabla \mathbf{u}_\mathbf{r} \right)^T \cdot \mathbf{S}_\mathbf{r}^\alpha - \mathbf{S}_\mathbf{r}^\alpha \cdot \left( \nabla \mathbf{u}_\mathbf{r} \right)$ is the upper-convected derivative of

$\mathbf{S}_\mathbf{r}^\alpha$, $D\mathbf{S}_\mathbf{r}^\alpha / Dt = \partial \mathbf{S}_\mathbf{r}^\alpha / \partial t + \mathbf{u}_\mathbf{r} \cdot \nabla \mathbf{S}_\mathbf{r}^\alpha$ is the material derivative, and the local mass-averaged velocity $\mathbf{u}_\mathbf{r}$ is calculated from Eq. (14) using the values of the free segment probabilities

$P_\mathbf{r}^\alpha$ and the momentum densities $\mathbf{g}_\mathbf{r}$ provided by the transport equations described below

and Eqs. (4)-(11). Here the dimensionless finite extensibility parameter [58]

$$b^\alpha = \frac{3\tau_Q^\alpha \left( N^\alpha \right)}{\tau_{db}^\alpha \left( N^\alpha \right)} = 3\left( N^\alpha - 1 \right) \tag{16}$$

is thrice the ratio between the relaxation time

$$\tau_Q^\alpha \left( N^\alpha \right) = \frac{N^\alpha \left( N^\alpha - 1 \right)^2 \tilde{a}_\alpha^2 \zeta_\mathbf{r}^\alpha}{24 k_B T} \qquad (17)$$

of a rigid dumbbell of extension $\left( N^\alpha - 1 \right) \tilde{a}_\alpha$, and the relaxation time

$$\tau_{db,\mathbf{r}}^\alpha = \frac{N^\alpha \left( N^\alpha - 1 \right) \tilde{a}_\alpha^2 \zeta_\mathbf{r}^\alpha}{24 k_B T} \qquad (18)$$

of a Hookean dumbbell with spring constant $3 k_B T / \left[ \left( N^\alpha - 1 \right) \tilde{a}_\alpha^2 \right]$. Note that $\tau_{db,\mathbf{r}}^\alpha$ is proportional to a local segmental friction coefficient $\zeta_\mathbf{r}^\alpha$. Note that in the limit $b^\alpha \to \infty$, Eq. (15) becomes the evolution equation for the second moment of the end-to-end distance of such a Hookean dumbbell.

At thermodynamic equilibrium, the second moment of the end-to-end distance $\mathbf{S}_\mathbf{r}^\alpha$ of a chain consisting of $N^\alpha$ Kuhn segments, joined together by $N^\alpha - 1$ freely jointed links, of length $a$ each, is a diagonal second-order tensor. The diagonal components are isotropic and have common value $\left( N^\alpha - 1 \right) a^2 / 3$. On the other hand, the equilibrium solution of Eq. (15) governing the evolution of $\mathbf{S}_\mathbf{r}^\alpha$ for the FENE-P model with the spring constant $3 k_B T \left[ \left( N^\alpha - 1 \right) \tilde{a}_\alpha^2 \left( 1 - \left\langle \left( \mathbf{Q}_\mathbf{r}^\alpha \right)^2 / Q_0^2 \right\rangle \right) \right]^{-1}$ yields

$$S_{\mathbf{r},xx}^\alpha = S_{\mathbf{r},yy}^\alpha = S_{\mathbf{r},zz}^\alpha = \left( \frac{b^\alpha}{b^\alpha + 3} \right) \frac{\left( N^\alpha - 1 \right) \tilde{a}_\alpha^2}{3}, \qquad S_{\mathbf{r},xy}^\alpha = S_{\mathbf{r},yz}^\alpha = S_{\mathbf{r},xz}^\alpha = 0, \qquad (19)$$

where $\tilde{a}_\alpha = c_\alpha a$. Thus Eq. (19) recovers the equilibrium value of $\mathbf{S}_\mathbf{r}^\alpha$ expected for a freely jointed chain of $N^\alpha$ Kuhn segments only if the FENE-P spring constant is tuned so that the relation $c_\alpha = \sqrt{\left( b^\alpha + 3 \right) / b^\alpha}$ is satisfied.

We proceed now to establish a relation between the values of $\mathbf{S}_{\mathbf{r}}^{\alpha}$ and the values of $\lambda_{\mathbf{a}_k,\mathbf{r}}^{\alpha}$ at any given moment in time. On the one hand, we know that in the FENE-P model, the probability distribution $\psi_{\mathbf{r}}^{\alpha}\left(\mathbf{Q}_{\mathbf{r}}^{\alpha}\right)$ for $\mathbf{Q}_{\mathbf{r}}^{\alpha}$, which is the solution of the appropriate Smoluchowski equation, has the form [6,58]

$$\psi_{\mathbf{r}}^{\alpha}\left(\mathbf{Q}_{\mathbf{r}}^{\alpha}\right) = \frac{\exp\left[-\frac{1}{2}\left(\mathbf{S}_{\mathbf{r}}^{\alpha}\right)^{-1}:\mathbf{Q}_{\mathbf{r}}^{\alpha}\mathbf{Q}_{\mathbf{r}}^{\alpha}\right]}{\sqrt{\left(2\pi\right)^{3}\det\left(\mathbf{S}_{\mathbf{r}}^{\alpha}\right)}}, \qquad (20)$$

where $\mathbf{S}_{\mathbf{r}}^{\alpha}$ is the solution of Eq. (15).

On the other hand, in the FCC lattice random walk representation of chains conformations of *ideal* (non-interacting) chains in a *homogeneous* fluid under flow, the time dependent probability distribution $P_{\mathbf{r}+\mathbf{Q}_{\mathbf{r}}^{\alpha},\mathbf{r}}^{\alpha}\left(N^{\alpha}-1\right)$ for a chain of $N^{\alpha}$ Kuhn segments of type $\alpha$ that starts at site $\mathbf{r}$ and ends at site $\mathbf{r}+\mathbf{Q}_{\mathbf{r}}^{\alpha}$ is derived recursively as follows. For ideal chains in a homogeneous bulk melt phase, the free segment statistical weights are constant: $\tilde{P}_{\mathbf{r}}^{\alpha} = c$ for all $\mathbf{r}$. Hence the recursion relation (7) assumes the form

$$P_{\mathbf{r}}^{\alpha}\left(s+1\right) = \sum_{k=1}^{12}\lambda_{-\mathbf{a}_k,\mathbf{r}+\mathbf{a}_k}^{\alpha}P_{\mathbf{r}+\mathbf{a}_k}^{\alpha}\left(s\right), \qquad (21)$$

where $P_{\mathbf{r}}^{\alpha}\left(s\right) = c\tilde{P}_{\mathbf{r}}^{\alpha}\left(s\right)$ for $s > 1$ and $P_{\mathbf{r}}^{\alpha}\left(1\right) = c$. Subtracting $P_{\mathbf{r}}^{\alpha}\left(s\right)$ from both sides of this equation, and using $\sum_{k=1}^{12}\lambda_{\mathbf{a}_k,\mathbf{r}}^{\alpha} = 1$, we get

$$P_{\mathbf{r}}^{\alpha}\left(s+1\right) - P_{\mathbf{r}}^{\alpha}\left(s\right) = \sum_{k=1}^{12}\left[\lambda_{-\mathbf{a}_k,\mathbf{r}}^{\alpha}P_{\mathbf{r}+\mathbf{a}_k}^{\alpha}\left(s\right) - \lambda_{\mathbf{a}_k,\mathbf{r}}^{\alpha}P_{\mathbf{r}}^{\alpha}\left(s\right)\right]. \qquad (22)$$

This equation is the master equation for the statistical weights of finding a terminus of ideal chains in a homogeneous melt under flow, generated as biased random walks on the FCC lattice. Using a Taylor series expansion for (22) and keeping terms to first order in $s$ and to second order in space, we see that $P_{\mathbf{r}}^{\alpha}(s)$ satisfies a diffusion equation

$$\frac{\partial P_{\mathbf{r}}^{\alpha}(s)}{\partial s} = \mathbf{\Lambda}_{\mathbf{r}}^{\alpha} : \nabla\nabla P_{\mathbf{r}}^{\alpha}(s), \qquad (23)$$

with the initial condition $P_{\mathbf{r}}^{\alpha}(1) = c$. We have assumed here that $\mathbf{\Lambda}_{\mathbf{r}'}^{\alpha} \approx \mathbf{\Lambda}_{\mathbf{r}}^{\alpha}$ for $|\mathbf{r}' - \mathbf{r}| < \kappa_{\mathbf{r}}^{\alpha}$ where $\kappa_{\mathbf{r}}^{\alpha}$ is the maximum eigenvalue of $\left\langle \mathbf{Q}_{\mathbf{r}}^{\alpha}\mathbf{Q}_{\mathbf{r}}^{\alpha} \right\rangle$ and used the symmetry $\lambda_{-\mathbf{a}_k,\mathbf{r}}^{\alpha} = \lambda_{\mathbf{a}_k,\mathbf{r}}^{\alpha}$ at each site. Collecting terms from the Taylor expansion, we get

$$\begin{aligned}
\Lambda_{xx,\mathbf{r}}^{\alpha} &= \frac{a^2}{2}\left[ \lambda_{\mathbf{a}_1,\mathbf{r}}^{\alpha} + \tfrac{1}{4}\left( \lambda_{\mathbf{a}_2,\mathbf{r}}^{\alpha} + \lambda_{\mathbf{a}_3,\mathbf{r}}^{\alpha} + \lambda_{\mathbf{a}_8,\mathbf{r}}^{\alpha} + \lambda_{\mathbf{a}_9,\mathbf{r}}^{\alpha} \right) \right] \\
\Lambda_{yy,\mathbf{r}}^{\alpha} &= \frac{a^2}{2}\left[ \tfrac{3}{4}\left( \lambda_{\mathbf{a}_2,\mathbf{r}}^{\alpha} + \lambda_{\mathbf{a}_3,\mathbf{r}}^{\alpha} \right) + \tfrac{1}{3}\lambda_{\mathbf{a}_7,\mathbf{r}}^{\alpha} + \tfrac{1}{12}\left( \lambda_{\mathbf{a}_8,\mathbf{r}}^{\alpha} + \lambda_{\mathbf{a}_9,\mathbf{r}}^{\alpha} \right) \right] \\
\Lambda_{zz,\mathbf{r}}^{\alpha} &= \frac{a^2}{2}\left[ \tfrac{2}{3}\left( \lambda_{\mathbf{a}_7,\mathbf{r}}^{\alpha} + \lambda_{\mathbf{a}_8,\mathbf{r}}^{\alpha} + \lambda_{\mathbf{a}_9,\mathbf{r}}^{\alpha} \right) \right] \\
\Lambda_{xy,\mathbf{r}}^{\alpha} &= \frac{a^2}{2}\left[ \tfrac{\sqrt{3}}{4}\left( \lambda_{\mathbf{a}_2,\mathbf{r}}^{\alpha} - \lambda_{\mathbf{a}_3,\mathbf{r}}^{\alpha} \right) + \tfrac{1}{4\sqrt{3}}\left( \lambda_{\mathbf{a}_8,\mathbf{r}}^{\alpha} - \lambda_{\mathbf{a}_9,\mathbf{r}}^{\alpha} \right) \right] \\
\Lambda_{yz,\mathbf{r}}^{\alpha} &= \frac{a^2}{2}\left[ \tfrac{\sqrt{2}}{6}\left( \lambda_{\mathbf{a}_8,\mathbf{r}}^{\alpha} + \lambda_{\mathbf{a}_9,\mathbf{r}}^{\alpha} - 2\lambda_{\mathbf{a}_7,\mathbf{r}}^{\alpha} \right) \right] \\
\Lambda_{xz,\mathbf{r}}^{\alpha} &= \frac{a^2}{2}\left[ \tfrac{1}{\sqrt{6}}\left( \lambda_{\mathbf{a}_8,\mathbf{r}}^{\alpha} - \lambda_{\mathbf{a}_9,\mathbf{r}}^{\alpha} \right) \right].
\end{aligned} \qquad (24)$$

The solution of Eq. (23) for $P_{\mathbf{r}+\mathbf{Q}_{\mathbf{r}}^{\alpha},\mathbf{r}}^{\alpha}\left( N^{\alpha}-1 \right)$ with the initial condition $P_{\mathbf{r},\mathbf{r}}^{\alpha}(1) = P_{\mathbf{r}}^{\alpha} = c$ is

$$P_{\mathbf{r}+\mathbf{Q}_{\mathbf{r}}^{\alpha},\mathbf{r}}^{\alpha}\left( N^{\alpha}-1 \right) = \frac{c\exp\left\{ -\tfrac{1}{2}\left[ 2\left( N^{\alpha}-1 \right)\mathbf{\Lambda}_{\mathbf{r}}^{\alpha} \right]^{-1} : \mathbf{Q}_{\mathbf{r}}^{\alpha}\mathbf{Q}_{\mathbf{r}}^{\alpha} \right\}}{\sqrt{\left(2\pi\right)^3 \det\left[ 2\left( N^{\alpha}-1 \right)\mathbf{\Lambda}_{\mathbf{r}}^{\alpha} \right]}}. \qquad (25)$$

By comparing $\tilde{P}_{\mathbf{r}+\mathbf{Q}_{\mathbf{r}}^{\alpha},\mathbf{r}}^{\alpha}\left( N^{\alpha}-1 \right) = P_{\mathbf{r}+\mathbf{Q}_{\mathbf{r}}^{\alpha},\mathbf{r}}^{\alpha}\left( N^{\alpha}-1 \right)/c$ obtained from (25) with (20), we get a system of six linear equations

$$\mathbf{S}_{\mathbf{r}}^{\alpha} = 2\left( N^{\alpha}-1 \right)\mathbf{\Lambda}_{\mathbf{r}}^{\alpha}. \qquad (26)$$

Using (24) to solve for the $\lambda^\alpha$'s, we get the solutions

$$\lambda_{\mathbf{a}_1,\mathbf{r}}^\alpha = \frac{1}{(N^\alpha-1)a^2}\left(\frac{1}{2}S_{xx,\mathbf{r}}^\alpha - \frac{1}{6}S_{yy,\mathbf{r}}^\alpha - \frac{1}{12}S_{zz,\mathbf{r}}^\alpha - \frac{1}{3\sqrt{2}}S_{yz,\mathbf{r}}^\alpha\right)$$

$$\lambda_{\mathbf{a}_2,\mathbf{r}}^\alpha = \frac{1}{(N^\alpha-1)a^2}\left(\frac{1}{3}S_{yy,\mathbf{r}}^\alpha - \frac{1}{12}S_{zz,\mathbf{r}}^\alpha + \frac{1}{\sqrt{3}}S_{xy,\mathbf{r}}^\alpha + \frac{1}{2\sqrt{6}}S_{yz,\mathbf{r}}^\alpha - \frac{1}{2\sqrt{6}}S_{xz,\mathbf{r}}^\alpha\right)$$

$$\lambda_{\mathbf{a}_3,\mathbf{r}}^\alpha = \frac{1}{(N^\alpha-1)a^2}\left(\frac{1}{3}S_{yy,\mathbf{r}}^\alpha - \frac{1}{12}S_{zz,\mathbf{r}}^\alpha - \frac{1}{\sqrt{3}}S_{xy,\mathbf{r}}^\alpha + \frac{1}{2\sqrt{6}}S_{yz,\mathbf{r}}^\alpha + \frac{1}{2\sqrt{6}}S_{xz,\mathbf{r}}^\alpha\right)$$

$$\lambda_{\mathbf{a}_7,\mathbf{r}}^\alpha = \frac{1}{(N^\alpha-1)a^2}\left(\frac{1}{4}S_{zz,\mathbf{r}}^\alpha - \frac{1}{\sqrt{2}}S_{yz,\mathbf{r}}^\alpha\right) \tag{27}$$

$$\lambda_{\mathbf{a}_8,\mathbf{r}}^\alpha = \frac{1}{(N^\alpha-1)a^2}\left(\frac{1}{4}S_{zz,\mathbf{r}}^\alpha + \frac{1}{2\sqrt{2}}S_{yz,\mathbf{r}}^\alpha + \frac{\sqrt{6}}{4}S_{xz,\mathbf{r}}^\alpha\right)$$

$$\lambda_{\mathbf{a}_9,\mathbf{r}}^\alpha = \frac{1}{(N^\alpha-1)a^2}\left(\frac{1}{4}S_{zz,\mathbf{r}}^\alpha + \frac{1}{2\sqrt{2}}S_{yz,\mathbf{r}}^\alpha - \frac{\sqrt{6}}{4}S_{xz,\mathbf{r}}^\alpha\right).$$

Substituting the equilibrium solution (19) into Eq. (27), we recover the static SCF values $\lambda_{\mathbf{a}_i,\mathbf{r}}^\alpha = \frac{1}{12}$.

### D. Self-Consistent Mean-Field Transport Theory

#### 1. Main Ideas and Relation to Existing Methodologies

At any given time, the local composition and kinematics in our inhomogeneous polymer fluid is described by the values of $\phi_\mathbf{r}^\alpha$ and $\mathbf{u}_\mathbf{r}^\alpha$, and the local rheology by the relation between the local deviatoric stress and the rate of strain $\nabla\mathbf{u}_\mathbf{r}$. The segmental volume fractions $\phi_\mathbf{r}^\alpha$ are determined from Eq. (8), using Eqs. (5), (7) and (9), provided that $P_\mathbf{r}^\alpha$ and $\lambda_{\mathbf{a}_1,\mathbf{r}}^\alpha$ are known. The latter are related by Eq. (27) to the conformation tensors $\mathbf{S}_\mathbf{r}^\alpha$ of ideal (noninteracting) chains. Their time evolution is described by Eq. (15), which depends on $\mathbf{u}_\mathbf{r}$. Hence if we know how to calculate the time evolution of $P_\mathbf{r}^\alpha$ and $\mathbf{u}_\mathbf{r}$ at each site $\mathbf{r}$, we are able to calculate the evolution of $\mathbf{S}_\mathbf{r}^\alpha$, $\lambda_{\mathbf{a}_1,\mathbf{r}}^\alpha$, and $\phi_\mathbf{r}^\alpha$ as well. Both $\mathbf{u}_\mathbf{r}$ and the stress are determined by a momentum balance equation expressing the conservation of linear momentum. Thus, in order to get a self-consistent, closed system of equations, it is necessary to augment the equations derived above with a set of time evolution equations for local free segment probabilities $P_{\mathbf{r}}^{\alpha}$ and momentum densities $\mathbf{g_r} = \mathbf{g_r}^A + \mathbf{g_r}^B$.

Before presenting a detailed derivation of the evolution equations for $P_{\mathbf{r}}^{\alpha}$ and $\mathbf{g_r}$, let us first explain the derivation method and discuss its relation to existing methodology. Our evolution equations for $P_{\mathbf{r}}^{\alpha}$ and $\mathbf{g_r}$ have the form of transport equations obtained by balancing fluxes of $P_{\mathbf{r}}^{\alpha}$ or $\mathbf{g_r}$, across the boundaries of a control volume of *segmental* dimensions, which is a unit cell of our FCC lattice. The flux expressions are derived using a mean-field approximation of joint probabilities by products of one-body probabilities and segmental volume fractions. This resembles derivation of transport equations arising from both the continuous [34] and lattice [35-39] versions of Boltzmann's kinetic theory for simple fluids [34], where the collision cross-section on a molecular scale determines the control volume. It bears even greater resemblance to the diffusive transport equations derived in kinetic mean-field theories [11] for stochastic lattice gases [59-61] from a microscopic master equation. Diffusive dynamics in stochastic lattice-gases and other Ising-like systems are modeled microscopically as thermally activated, stochastic pair exchange events between adjacent sites [59,60]. The exchange occurs only if the adjacent sites are occupied by a molecule and a vacancy in a lattice gas model, or by opposite spins in an Ising system. The activated pair exchange events occur with certain transition rates depending on the change in the nonequilibrium free energy caused by the pair exchange, in dimensionless units set by the temperature. To assure convergence to thermodynamic equilibrium, the transition rates have to satisfy a local detailed balance condition. The time scale for the transition rate is set by the appropriate diffusion constant. The time evolution of the system's configurational probability is a Markov process governed by a master equation [59].

The complexity of the problem precludes an analytic solution to the master equation, except for very simple limiting cases. Such Markov processes can be simulated directly using Monte Carlo (MC) methods [62], which do not neglect many-body correlations, but have strong local fluctuations that may blur the connection to mesoscopic descriptions, such as the Cahn-Hilliard-Cook [41,63], time-dependent Landau-Ginzburg [21,42], and phase-field models [64,65]. This is avoided by the kinetic mean-field theories [11] that were developed for stochastic lattice-gas models [66,67]. Similarly to equilibrium mean-field theories, kinetic mean-field theories approximate the configurational joint probability for system constituents by a product of one-body probabilities for each microscopic constituent, interacting solely with an external potential field. The interaction of a constituent with its conjugate field is expressed as a mean configurational energy of interaction with all the other system constituents. However, in *kinetic* mean-field theories both the mean fields and the one-body probability distributions are *time-dependent*. The time evolution of the local one-body probabilities and of the self-consistent fields is obtained by balancing fluxes of one-body probabilities about a control volume on molecular length scale. The resulting transport equations are formulated on that scale. The factorization of joint probabilities into a product of one-body probabilities suppresses correlations and fluctuations in these models, making them akin to mesoscopic, deterministic transport equations, such as Cahn-Hilliard-Cook [41,63] and time-dependent Landau-Ginzburg [21,42] equations, typically described by high-order, nonlinear PDEs. They are very different from classical macroscopic equations, such as the diffusion or Fokker-Plank equations, which are typically linear, second-order PDEs with Gaussian solutions describing fluctuations about a deterministically evolving mean. The latter can be formally obtained as a low-order truncation of the so called $\Omega$-expansion of the master equation in powers of $\Omega^{(1-n)/2}$ developed by van Kampen [44], where $\Omega$ is associated with the number of microscopic constituents in a minimal control volume. Low order truncation of the $\Omega$-expansion is justified only for large $\Omega$, and, as noted by van Kampen [44], keeping successively higher order terms in the $\Omega$-expansion gives rise to higher order derivatives and nonlinear coupling in the resulting PDEs for the continuous probability density, which are responsible for the deviation of fluctuations from a Gaussian distribution. Indeed, when binning averages are performed on trajectories of MD simulations using bins with spatial dimensions approaching the dimensions of a single constituent, the resulting averages typically exhibit strong fluctuations which cannot be described as small Gaussian fluctuations about a deterministically evolving mean.

A dynamic SCF lattice gas model for simple fluids has been derived by Khan and Shnidman [3,4], and used for molecular-scale computational modeling of interfacial and wetting flows. Their model includes convective contributions to one-body probability fluxes due to cage advection, as well as a diffusive contribution to the one-body probability fluxes arising from activated hopping between cages as described above. Momentum fluxes similarly include a convective contribution due to cage advection, a viscous contribution from momentum transferred by activated hops, as well as a term representing friction forces generated by net cage velocities arising from biased hopping across bonds. This dynamic SCF model for simple fluids has been used successfully to study both isothermal and nonisothermal interfacial and wetting dynamics in one, two and three-component compressible simple fluids. Truncated Taylor expansions of the evolution equations of this DSCF lattice model lead to continuous evolution equations similar to those of Model H [21,42], a time-dependent Landau Ginzburg model coupling convective and diffusive transport of local mean species densities with convective and viscous transport of momentum.

Note that the PDEs describing the continuous phenomenological models above indeed involve higher (fourth) order derivative terms and nonlinearities which are crucial for modeling interfacial dynamics. Our transport equations have the form of a system of nonlinear ODEs. Using truncated Taylor expansions, they can be approximated by a system of nonlinear, higher-order PDEs similar in form to the continuous, deterministic transport models formulated phenomenologically on the mesoscopic scale. However, not only the solution of high order, nonlinear PDEs is harder than solving a system of nonlinear ODEs provided by our model, but truncation of the $\Omega$ expansions at higher orders leads to unphysical behavior, such as violation of the positivity of the probability, and of its approach to a steady state.

A central objective of our polymer DSCF theory is to extend the approach described above to unentangled polymer fluids. This requires accounting for intrachain correlations and for deformation of chain conformations under flow.

## 2. Time Evolution of Free Segment Probabilities

The balance of fluxes of one-body probabilities $P_{\mathbf{r}}^{\alpha}$ of species $\alpha$ across a unit cell centered at site $\mathbf{r}$ leads to the following time evolution equation for the free segment probability density of species $\alpha$ within the control volume of a unit cell,

$$\frac{dP_{\mathbf{r}}^{\alpha}}{dt} = -\nabla \cdot \left( P_{\mathbf{r}}^{\alpha} \mathbf{u}_{\mathbf{r}} \right) - \sum_{k=1}^{12} \frac{j_{\mathbf{r},\mathbf{r}+\mathbf{a}_k}^{\alpha}}{a} \ . \qquad (28)$$

The first term on the RHS of Eq. (28) is the divergence of convective fluxes of free segment probabilities. In our computations we use a second-order centered difference approximation for the FCC lattice, except at the first and the last layers adjacent to the bottom and top walls, for which the second-order forward and the backward difference approximations are used, respectively. The second term on the RHS of (28) represents the divergence of diffusive fluxes of the free-segment probability $P_{\mathbf{r}}^{\alpha}$ due to hops into and out of the control volume from/into the adjacent sites, as shown in Fig. 4 in the case of a triangular lattice, for simplicity. The net diffusive flux of free segments of species $\alpha$ across each bond is given by

$$
\begin{aligned}
\mathbf{j}_{\mathbf{r},\mathbf{r}+\mathbf{a}_k}^{\alpha} = & \left(1 - \delta_{(\mathbf{r}+\mathbf{a}_k)\cdot\mathbf{e}_3, 0}\right)\left(1 - \delta_{(\mathbf{r}+\mathbf{a}_k)\cdot\mathbf{e}_3, z_L+\sqrt{2/3}a}\right)\left[\frac{D_{\mathbf{r},\mathbf{r}+\mathbf{a}_k}^{\alpha}}{\left(1-\overline{\phi}\right)}\right] \\
& \times \left[P_{\mathbf{r}}^{\alpha}\left(1-\phi_{\mathbf{r}+\mathbf{a}_k}\right)\varphi\left(\frac{\Delta_{\mathbf{r},\mathbf{r}+\mathbf{a}_k}\left\langle\mathcal{H}_{\mathbf{r}}^{\alpha}\right\rangle}{k_B T}\right) - P_{\mathbf{r}+\mathbf{a}_k}^{\alpha}\left(1-\phi_{\mathbf{r}}\right)\varphi\left(-\frac{\Delta_{\mathbf{r},\mathbf{r}+\mathbf{a}_k}\left\langle\mathcal{H}_{\mathbf{r}}^{\alpha}\right\rangle}{k_B T}\right)\right]\frac{\mathbf{a}_k}{a}.
\end{aligned}
\tag{29}
$$

It is assumed that there are no fluxes coming into or out of the walls. The factors $\left(1 - \delta_{(\mathbf{r}+\mathbf{a}_k)\cdot\mathbf{e}_3, 0}\right)$ and $\left(1 - \delta_{(\mathbf{r}+\mathbf{a}_k)\cdot\mathbf{e}_3, z_L+\sqrt{2/3}a}\right)$ suppress any contributions to fluxes from bottom and top walls, respectively. The net flux across a bond consists of the finite difference gradient of the probability current due to activated hops between the two adjacent sites. Consider a Markov chain model for the hopping flux from site $\mathbf{r}$ to site $\mathbf{r}+\mathbf{a}_k$. It is given by the product of the initial probability $P_{\mathbf{r}}^{\alpha}$ for the site $\mathbf{r}$ to be occupied by a free segment and the transition rate $\left[D_{\mathbf{r},\mathbf{r}+\mathbf{a}_k}^{\alpha}/\left(1-\overline{\phi}\right)\right]\left(1-\phi_{\mathbf{r}+\mathbf{a}_k}\right)\varphi\left(\Delta_{\mathbf{r},\mathbf{r}+\mathbf{a}_k}\left\langle\mathcal{H}_{\mathbf{r}}^{\alpha}\right\rangle/k_B T\right)$ to hop from $\mathbf{r}$ to $\mathbf{r}+\mathbf{a}_k$. This transition rate should vanish if the site $\mathbf{r}+\mathbf{a}_k$ is not vacant. This is assured by the factor $\left(1-\phi_{\mathbf{r}+\mathbf{a}_k}\right)$ which is the probability that the site $\mathbf{r}+\mathbf{a}_k$ is not occupied by a segment, including intrachain correlations in a chain of $N^\alpha$ freely jointed Kuhn segments. As discussed before, intrachain correlation effects are pronounced near walls and interfaces, where chain conformations are constrained. Note that $\phi_{\mathbf{r}}^\alpha = P_{\mathbf{r}}^\alpha$ in homogeneous or simple (monomer) fluids. For simple fluids [3,4], this form of diffusive fluxes is identical to the diffusive fluxes used in kinetic mean-field theories [11] modeling diffusive hopping transport in stochastic lattice gases [59-61,68]. Since hopping is an activated rate process, the transition rate is proportional to $\varphi\left(\Delta_{\mathbf{r},\mathbf{r}+\mathbf{a}_k}\left\langle\mathcal{H}_{\mathbf{r}}^\alpha\right\rangle / k_B T\right)$, where the form of the rate function $\varphi$ realizes a particular coupling to a heat bath. As a necessary condition to achieve convergence of the dynamic model to an equilibrium state under equilibrium conditions, the rate function has to satisfy the so-called local detailed balance relation [61]

$$\varphi(h) = e^{-h}\varphi(-h).$$ (30)

In our computation we use the Kawasaki form for the rate function (see discussion of this choice below)

$$\varphi(h) = \frac{2e^{-h/2}}{e^{h/2} + e^{-h/2}}.$$ (31)

This form of the transition rate function is certainly not unique [11,61]. There are many possible functional forms of the transition rate function satisfying Eq. (30), as there are many different possible realizations of the couplings to the heat bath in real systems as well. This is a well-known problem, which is by no means limited to the particular method for modeling nonequilibrium dynamics chosen here. For example, in the Langevin approach a particular coupling to the heat bath is realized by the functional form of the fluctuating noise term. As a necessary condition for converging to an equilibrium state under equilibrium conditions, the noise term has to satisfy the fluctuation dissipation theorem, but there are many different forms of the noise term satisfying. Similarly, there is a multitude of different methods for realizing the coupling to the heat bath in MD simulations, which have to satisfy some necessary conditions to reproduce equilibrium statistics under equilibrium conditions, but may lead to differences in nonequilibrium evolution.

In practice, it is found that different realizations of the coupling to the heat bath may certainly affect the short-time and small-scale details of the dynamics, but recover the same qualitative trends at longer times and large scales. Hence the choice of a particular realization of the coupling to the heat bath in a nonequilibrium model is guided by simplifying assumptions idealizing such couplings in real systems, in order to make modeling practical. For example, in the Langevin approach a $\delta$-function form for the "white" noise is commonly used for analytical calculations, while a finite Gaussian form may be used in simulations. For kinetic lattice gases, the simplest analytical form for the transition rate function that satisfies (30) is $\varphi(h) = e^{-h/2}$. However, since it is not bounded in the limit $h \to -\infty$, it may cause frequent crashes in computer simulations due to overflows. The Metropolis form $\varphi(h) = \min\left\{1, e^{-h/2}\right\}$ is bounded, and indeed has been used extensively in Monte-Carlo simulations, but its derivative has a discontinuity at $h = 0$. This may not be important in Monte-Carlo simulations, where the discontinuity is smeared by fluctuations, but it may cause stability problems when solving the mean-field ordinary differential equations. The Kawasaki form of the rate function in Eq. (31) satisfies the local detailed balance relation (30), is bounded between 0 and 2, and is a smooth function.

The argument of the transition rate function $\Delta_{\mathbf{r}, \mathbf{r}+\mathbf{a}_k} \left\langle \mathcal{H}_{\mathbf{r}}^{\alpha} \right\rangle / k_B T$ represents the change in the site contributions $\left\langle \mathcal{H}_{\mathbf{r}}^{\alpha} \right\rangle$ to the self-consistent free energy that is caused by a hop of a single $\alpha$-type Kuhn segment from site $\mathbf{r}$ to an adjacent vacant site $\mathbf{r}+\mathbf{a}_k$. Note that in a homogeneous equilibrium system the quantities $P_{\mathbf{r}}^{\alpha}$, $\phi_{\mathbf{r}}^{\alpha}$, $\left\langle \mathcal{H}_{\mathbf{r}}^{\alpha} \right\rangle$ and $D_{\mathbf{r}, \mathbf{r}+\mathbf{a}_k}^{\alpha}$ are independent of $\mathbf{r}$ and $\mathbf{a}_k$, and $\Delta_{\mathbf{r}, \mathbf{r}+\mathbf{a}_k} \left\langle \mathcal{H}_{\mathbf{r}}^{\alpha} \right\rangle = 0$. Therefore the net diffusive currents across each bond vanish in such a system. We also know that $\phi_{\mathbf{r}}^{\alpha}$, $\left\langle \mathcal{H}_{\mathbf{r}}^{\alpha} \right\rangle$, and $D_{\mathbf{r}, \mathbf{r}+\mathbf{a}_k}^{\alpha}$ are nonlinear functions of $P_{\mathbf{r}}^{\alpha}$. Consider small composition perturbations from an equilibrium homogeneous state far from the wall, while constraining the velocity field $\mathbf{u}_{\mathbf{r}}$ to be uniform. Noting that $1 - \phi_{\mathbf{r}} = 1 - \bar{\phi}$ and $\varphi(0) = 1$ in Eq. (31), Eq. (29) assumes the form

$$\mathbf{j}_{\mathbf{r}, \mathbf{r}+\mathbf{a}_k}^{\alpha} = -D^{\alpha} \left( \frac{P_{\mathbf{r}+\mathbf{a}_k}^{\alpha} - P_{\mathbf{r}}^{\alpha}}{a} \right) \mathbf{a}_k, \tag{32}$$

which is similar to Fick's First Law. Since the time evolution of $\phi_{\mathbf{r}}^{\alpha}$ is determined by the time evolution of $P_{\mathbf{r}}^{\alpha}$, we identify $D^{\alpha}$ with the translational self diffusion coefficient in the homogeneous equilibrium system, given by the Stokes-Einstein relation

$$D^{\alpha} = \frac{k_B T}{N^{\alpha} \zeta^{\alpha}}, \tag{33}$$

where $N^{\alpha} \zeta^{\alpha}$ is chain friction coefficient, and $\zeta^{\alpha}$ is the segmental friction coefficient. For an inhomogeneous nonequilibrium system we assume

$$D_{\mathbf{r}, \mathbf{r}\pm\mathbf{a}_k}^{\alpha} = \frac{k_B T}{N^{\alpha} \zeta_{\mathbf{r}, \mathbf{r}\pm\mathbf{a}_k}^{\alpha}}, \tag{34}$$

where $\zeta_{\mathbf{r},\mathbf{r}\pm\mathbf{a}_k}^{\alpha} = \sqrt{\zeta_{\mathbf{r}}^{\alpha}\zeta_{\mathbf{r}\pm\mathbf{a}_k}^{\alpha}}$ is the geometric average of the local segmental friction coefficients at the two sites $\mathbf{r}$ and $\mathbf{r}+\mathbf{a}_k$. For a polymer melt above its glass transition, it is well known that local friction coefficients have a strong dependence on the local free volume $1-\phi_{\mathbf{r}}$. This dependence is captured by the Doolittles' law [69]

$$\zeta_{\mathbf{r}}^{\alpha} = \zeta_0^{\alpha} \exp\left\{ \left(1-\phi_{\mathbf{r}}\right)^{-1} - \left(1-\overline{\phi}\right) \right\}, \qquad (35)$$

where $\zeta_0^{\alpha}$ is the friction coefficient in a homogeneous melt at the reference temperature and pressure, and whose density is $\overline{\phi} = \overline{\phi}\left(T_{ref}, P_{ref}\right)$. Note that Eq. (35) recovers Doolittles' law for the free volume relaxation time, $\tau_f = \tau_0 \, \mathrm{e}^{1/f}$ in a system with homogeneous distribution of free volume $f = 1 - \overline{\phi}$. Doolittles' law is equivalent to the Vogel-Fulcher-Tammann-Hesse relation [70-73] and to the Williams-Landel-Ferry equation [74,75], and thus ensures recovery of the time-temperature superposition in a homogeneous system.

Within the framework of a dynamic self-consistent mean-field approximation described above, if a Kuhn segment of type $\alpha$ occupies a site $\mathbf{r}$, it makes the following contribution to the generalized (nonequilibrium) free energy:

$$\frac{\left\langle \mathcal{H}_{\mathbf{r}}^{\alpha} \right\rangle}{k_B T} = \frac{1}{2}\sum_{\beta}\left(1-\delta_{\alpha\beta}\right)\chi_{\alpha\beta}\left\langle\left\langle \phi_{\mathbf{r}}^{\beta} \right\rangle\right\rangle - \chi_s^{\alpha}\left(\delta_{z,z_1} + \delta_{z,z_L}\right) + \frac{m^{\alpha}}{2k_B T}\phi_{\mathbf{r}}^{\alpha}\mathbf{u}_{\mathbf{r}}^2 + \frac{f_{\mathbf{r}}^{\alpha}w}{k_B T}. \qquad (36)$$

The first term on RHS of the equation above accounts for interactions between a pair of nearest-neighbor segments. These interactions are characterized by the segment-segment interaction parameter, $\chi_{\alpha\beta}$, which is related to the Flory-Huggins [26] interaction parameter $12\chi_{\alpha\beta}$, defined as the energy change (normalized by $k_B T$) due to the transfer of an $\alpha$ segment from a solution of pure $\alpha$ to a solution of pure $\beta$. For segments of the same size, it is assumed that $\chi_{AA} = \chi_{BB} = 0$ and that $\chi_{AB} = \chi_{BA}$. Within a mean-field approximation [31-33], the form of the segment-segment interaction energy is obtained by averaging over all contacts that an $\alpha$-type segment, located at site $\mathbf{r}$, has with segments of type $\beta$, located at sites $\mathbf{r} + \mathbf{a}_k$. We use the double angular brackets to denote summation over all nearest-neighbors. Therefore, for the FCC lattice between the two walls we have

$$\left\langle\!\left\langle \phi_{\mathbf{r}}^{\beta} \right\rangle\!\right\rangle = \sum_{k=1}^{6} \phi_{\mathbf{r}+\mathbf{a}_k}^{\beta} + \sum_{k=7}^{9} \left(1 - \delta_{z,z_L}\right) \phi_{\mathbf{r}+\mathbf{a}_k}^{\beta} + \sum_{k=10}^{12} \left(1 - \delta_{z,z_1}\right) \phi_{\mathbf{r}+\mathbf{a}_k}^{\beta} . \tag{37}$$

Note that, since the segments at adjacent sites belong to jointed chains of Kuhn segments, the probabilistic averaging over segmental occupancies at the adjacent sites uses segmental volume fractions at these sites, rather than free segment probabilities, to account for intrachain correlations.

The second term on the right-hand side of (36) represents interactions between the segments in the first and the last layers and the solid walls. The wall interaction parameter $\chi_s^{\alpha}$, is defined as the energy change (normalized by $k_B T$) due to the transfer in a pure-$\alpha$ fluid of an $\alpha$ segment to a layer adjacent to a wall from a layer further away from the wall. It is positive for attractive segment-wall interactions, and negative for repulsive segment-wall interactions.

The first two terms are the familiar segmental interaction terms, that have the same form as in the equilibrium lattice SCF theory of Scheutjens and Fleer [12]. The last two terms make a nontrivial (i.e. not a constant) contribution to the generalized free energy only if the system is driven out of thermodynamic equilibrium by applied stresses, establishing a nonuniform velocity flow field $\mathbf{u_r}$, which, in turn, perturbs the conformation tensor $\mathbf{S_r^\alpha}$ corresponding to the second moment of the en-to-end distance of ideal jointed chains from its equilibrium isotropic form. The third term is the kinetic energy contribution arising from the local mean convective velocity of segments advected by the flow, based on the local equilibrium approximation defined in Eq. (3). A similar term arises in a dynamic self-consistent mean-field theory for simple fluids [3,4]. The last term is not present in simple (monomer) fluids. It accounts for the local contribution to the free energy due to the deformation of polymer chains by the nonuniform flow, and is given by [10]

$$f_{\mathbf{r}}^\alpha = \tilde{u}_{\mathbf{r}} - T\tilde{s}_{\mathbf{r}} = -n_{\mathbf{r}}^\alpha k_B T \int \psi_{\mathbf{r}}^\alpha \left[ \ln \psi_{0,\mathbf{r}}^\alpha - \ln \psi_{\mathbf{r}}^\alpha \right] d\mathbf{Q}_{\mathbf{r}}^\alpha . \qquad (38)$$

Here $n_{\mathbf{r}}^\alpha = \left( \phi_{\mathbf{r}}^\alpha w^{-1} \right) \big/ N^\alpha$ is the number of chains composed of segments of type $\alpha$ per unit volume, $\tilde{u}_{\mathbf{r}}$ is the local internal energy of the polymer chains per unit volume, defined by the elastic potential energy of the springs in the FENE-P model, $\tilde{s}_{\mathbf{r}}$ is the configurational contribution of the chains to the entropy per unit volume, $\psi_{\mathbf{r}}^\alpha \left( \mathbf{Q}_{\mathbf{r}}^\alpha \right)$ is the local time-dependent probability distribution for finding a chain of type $\alpha$ with the center-of-mass at site $\mathbf{r}$ and the end-to-end distance $\mathbf{Q}_{\mathbf{r}}^\alpha$, given by Eq. (20), and $\psi_{\mathbf{r},0}^\alpha \left( \mathbf{Q}_{\mathbf{r}}^\alpha \right)$ is $\psi_{\mathbf{r}}^\alpha \left( \mathbf{Q}_{\mathbf{r}}^\alpha \right)$ at equilibrium. Evaluating the integral in Eq. (38) we get [10]

$$\frac{f_{\mathbf{r}}^\alpha}{k_B T} = -\frac{1}{2} n_{\mathbf{r}}^\alpha \left\{ \mathrm{tr}\left( \boldsymbol{\delta}_{\mathbf{r}} - \frac{3}{\left( N^\alpha - 1 \right) \tilde{a}_\alpha^2} \mathbf{S}_{\mathbf{r}}^\alpha \right) + \ln\left[ \det\left( \frac{3}{\left( N^\alpha - 1 \right) \tilde{a}_\alpha^2} \mathbf{S}_{\mathbf{r}}^\alpha \right) \right] \right\} . \qquad (39)$$

### 3. Time Evolution of Momentum Density

Let us consider first the time evolution of momentum density in an incompressible system of ideal (non-interacting) chains of type $\alpha$ that are modeled as FENE-P dumbbells. The time evolution of the chain conformation tensor is given by Eq. (15), which we rewrite as follows,

$$\overset{\triangledown}{\mathbf{S}_{\mathbf{r}}^{\alpha}} = -\frac{1}{\tau_{db,\mathbf{r}}^{\alpha}}\left[\frac{\mathbf{S}_{\mathbf{r}}^{\alpha}}{M_{\mathbf{r}}^{\alpha}} - \frac{\left(N^{\alpha}-1\right)\tilde{a}_{\alpha}^{2}}{3}\boldsymbol{\delta}\right], \qquad (40)$$

where

$$M_{\mathbf{r}}^{\alpha} = 1 - \frac{3}{\left(N^{\alpha}-1\right)\tilde{a}_{\alpha}^{2}b^{\alpha}}\mathrm{Tr}\mathbf{S}_{\mathbf{r}}^{\alpha} \qquad (41)$$

originates from the nonlinear spring force in the FENE-P model ($M_{\mathbf{r}}^{\alpha} = 1$ in Hookean dumbbell model that is recovered in the limit $b^{\alpha} \to \infty$) and $\tau_{db,\mathbf{r}}^{\alpha}$ is the local relaxation time for a Hookean dumbbell. Multiplication of Eq. (40) by $\tilde{\tau}_{db,\mathbf{r}}^{\alpha} = \tau_{db,\mathbf{r}}^{\alpha}M_{\mathbf{r}}^{\alpha}$ and rearrangement of the resulting equation gives

$$\tilde{\tau}_{db,\mathbf{r}}^{\alpha}\overset{\triangledown}{\mathbf{S}_{\mathbf{r}}^{\alpha}} + \mathbf{S}_{\mathbf{r}}^{\alpha} = \frac{\left(N^{\alpha}-1\right)\tilde{a}_{\alpha}^{2}}{3}M_{\mathbf{r}}^{\alpha}\boldsymbol{\delta}. \qquad (42)$$

Let us define a *deviatoric* chain conformation tensor as follows

$$\mathbf{T}_{\mathbf{r}}^{\alpha} = \mathbf{S}_{\mathbf{r}}^{\alpha} - \frac{\left(N^{\alpha}-1\right)\tilde{a}_{\alpha}^{2}}{3}M_{\mathbf{r}}^{\alpha}\boldsymbol{\delta}. \qquad (43)$$

Using (43) to get $\mathbf{S}_{\mathbf{r}}^{\alpha}$ as a function of $\mathbf{T}_{\mathbf{r}}^{\alpha}$, and substituting it into Eq. (42), we obtain

$$\tilde{\tau}_{db,\mathbf{r}}^{\alpha}\left(\overset{\triangledown}{\mathbf{T}_{\mathbf{r}}^{\alpha}} + \frac{\left(N^{\alpha}-1\right)\tilde{a}_{\alpha}^{2}}{3}\overset{\triangledown}{\overbrace{\left(M_{\mathbf{r}}^{\alpha}\boldsymbol{\delta}\right)}}\right) + \mathbf{T}_{\mathbf{r}}^{\alpha} = 0, \quad (44)$$

where

$$\overset{\triangledown}{\left(M_{\mathbf{r}}^{\alpha}\boldsymbol{\delta}\right)} = M_{\mathbf{r}}^{\alpha}\overset{\triangledown}{\boldsymbol{\delta}} + \frac{\partial M_{\mathbf{r}}^{\alpha}}{\partial t}\boldsymbol{\delta} + \left(\mathbf{u_r}\cdot\nabla M_{\mathbf{r}}^{\alpha}\right)\boldsymbol{\delta}. \quad (45)$$

The upper-convective time derivative of the unit tensor is given by

$$\overset{\triangledown}{\boldsymbol{\delta}} = \dot{\boldsymbol{\delta}} - \nabla\mathbf{u_r}^{\mathrm{T}}\cdot\boldsymbol{\delta} - \boldsymbol{\delta}\cdot\nabla\mathbf{u_r} = -2\mathbf{D_r}, \qquad (46)$$

where

$$\mathbf{D_r} = \frac{1}{2}\left(\nabla\mathbf{u_r} + \nabla\mathbf{u_r}^{\mathrm{T}}\right) \qquad (47)$$

is the rate-of-deformation tensor (i.e. the symmetric part of the velocity gradient). Substituting Eq. (45) into Eq. (44) and using Eq. (46) yields the following expression for the deviatoric chain conformation tensor

$$\begin{aligned}
\mathbf{T_r}^{\alpha} = {}& 2\tilde{\tau}_{db,\mathbf{r}}^{\alpha}\frac{\left(N^{\alpha}-1\right)\tilde{a}_{\alpha}^{2}}{3}M_{\mathbf{r}}^{\alpha}\mathbf{D_r} \\
& - \tilde{\tau}_{db,\mathbf{r}}^{\alpha}\left[\overset{\triangledown}{\mathbf{T_r}^{\alpha}} + \frac{\left(N^{\alpha}-1\right)\tilde{a}_{\alpha}^{2}}{3}\left(\frac{\partial M_{\mathbf{r}}^{\alpha}}{\partial t}\boldsymbol{\delta} + \left(\mathbf{u_r}\cdot\nabla M_{\mathbf{r}}^{\alpha}\right)\boldsymbol{\delta}\right)\right],
\end{aligned} \qquad (48)$$

If $n_{\mathbf{r}}^{\alpha}$ is the local density of ideal (non-interacting) FENE-P chains of type $\alpha$ at site $\mathbf{r}$, their contribution to the stress is

$$\boldsymbol{\sigma}_{\mathbf{r}}^{\alpha} = n_{\mathbf{r}}^{\alpha}\frac{3k_{B}T}{M_{\mathbf{r}}^{\alpha}\left(N^{\alpha}-1\right)\tilde{a}_{\alpha}^{2}}\mathbf{S_r}^{\alpha}, \qquad (49)$$

where the chain density

$$n_{\mathbf{r}}^{\alpha} = \frac{\phi_{\mathbf{r}}^{\alpha}}{wN_{\mathbf{r}}^{\alpha}} \qquad (50)$$

is constant in an incompressible system. The contribution of these chains to the *deviatoric* stress tensor at site $\mathbf{r}$ is defined as

$$\boldsymbol{\tau}_{\mathbf{r}}^{\alpha} = n_{\mathbf{r}}^{\alpha}\frac{3k_{B}T}{M_{\mathbf{r}}^{\alpha}\left(N^{\alpha}-1\right)\tilde{a}^{2}}\mathbf{T_r}^{\alpha}. \qquad (51)$$

Multiplying Eq. (48) by $n_{\mathbf{r}}^{\alpha}\left[3k_BT/M_{\mathbf{r}}^{\alpha}\left(N^{\alpha}-1\right)\tilde{a}_{\alpha}^2\right]$, and using $\tau_{db,\mathbf{r}}^{\alpha}=\tilde{\tau}_{db,\mathbf{r}}^{\alpha}/M_{\mathbf{r}}^{\alpha}$ and Eq. (18) defining the Hookean relaxation time $\tau_{db,\mathbf{r}}^{\alpha}$, we obtain

$$\boldsymbol{\tau}_{\mathbf{r}}^{\alpha}=2\eta_{\mathbf{r}}^{\alpha}\mathbf{D}_{\mathbf{r}}-\boldsymbol{\varepsilon}_{\mathbf{r}}^{\alpha}\,, \qquad (52)$$

where

$$\boldsymbol{\varepsilon}_{\mathbf{r}}^{\alpha}=\frac{n_{\mathbf{r}}^{\alpha}N^{\alpha}\zeta_{\mathbf{r}}^{\alpha}}{8}\left[\overset{\triangledown}{\mathbf{T}_{\mathbf{r}}^{\alpha}}+\frac{\left(N^{\alpha}-1\right)\tilde{a}_{\alpha}^2}{3}\left(\frac{\partial M_{\mathbf{r}}^{\alpha}}{\partial t}\boldsymbol{\delta}+\left(\mathbf{u}_{\mathbf{r}}\cdot\nabla M_{\mathbf{r}}^{\alpha}\right)\boldsymbol{\delta}\right)\right]\,, \qquad (53)$$

and both $n_{\mathbf{r}}^{\alpha}$ and $\zeta_{\mathbf{r}}^{\alpha}$ are constant in an incompressible system. In Eq. (52),

$$\eta_{\mathbf{r}}^{\alpha}=\tilde{\tau}_{db,\mathbf{r}}^{\alpha}G_{\mathbf{r}}^{\alpha}=n_{\mathbf{r}}^{\alpha}M_{\mathbf{r}}^{\alpha}N^{\alpha}\left(N^{\alpha}-1\right)\zeta_{\mathbf{r}}^{\alpha}\tilde{a}_{\alpha}^2/24 \qquad (54)$$

is the contribution to the local shear viscosity in a system of ideal chains of type $\alpha$ with local density $n_{\mathbf{r}}^{\alpha}$. Here $G_{\mathbf{r}}^{\alpha}=n_{\mathbf{r}}^{\alpha}k_BT$ is the chain contribution to the bulk modulus and $\tilde{\tau}_{db,\mathbf{r}}^{\alpha}$ is the effective relaxation time for FENE-P dumbbells. Thus the first term on RHS of Eq. (52) is a Newtonian viscous contribution to the deviatoric stress. The second term on RHS of Eq. (52) vanishes for a homogeneous system at the steady state. We will refer to it as the *elastic* contribution to the *deviatoric* stress. The time evolution of the momentum density of segments belonging to a melt of ideal chains of type $\alpha$ is given by

$$\frac{d\mathbf{g}_{\mathbf{r}}^{\alpha}}{dt}=\nabla\cdot\left(-\mathbf{g}_{\mathbf{r}}^{\alpha}\mathbf{u}_{\mathbf{r}}-\boldsymbol{\varepsilon}_{\mathbf{r}}^{\alpha}+2\eta_{\mathbf{r}}^{\alpha}\mathbf{D}_{\mathbf{r}}\right)\,, \qquad (55)$$

where $-\mathbf{g}_{\mathbf{r}}^{\alpha}\mathbf{u}_{\mathbf{r}}$, $-\boldsymbol{\varepsilon}_{\mathbf{r}}^{\alpha}$ and $2\eta_{\mathbf{r}}^{\alpha}\mathbf{D}_{\mathbf{r}}$ are the convective, elastic and viscous contributions to the stress, respectively. Note that for a one-component incompressible fluid, where $\rho=\rho_{\mathbf{r}}^{\alpha}$ and $\eta_{\mathbf{r}}^{\alpha}$ are constant, $\nabla\cdot\mathbf{u}_{\mathbf{r}}=0$, and $\mathbf{g}_{\mathbf{r}}^{\alpha}=\rho_{\mathbf{r}}^{\alpha}\mathbf{u}_{\mathbf{r}}^{\alpha}=\rho_{\mathbf{r}}^{\alpha}\mathbf{u}_{\mathbf{r}}$. The viscous term represents diffusion of the momentum density components

$$\nabla \cdot \left(2\eta_{\mathbf{r}}^{\alpha}\mathbf{D_r}\right) = \eta_{\mathbf{r}}^{\alpha}\nabla \cdot \left(\nabla\mathbf{u_r} + \nabla\mathbf{u_r}^{T}\right) = \nu_{\mathbf{r}}^{\alpha}\nabla^{2}\mathbf{g_r}^{\alpha} , \tag{56}$$

where the kinematic viscosity $\nu_{\mathbf{r}}^{\alpha} = \eta_{\mathbf{r}}^{\alpha}/\rho_{\mathbf{r}}^{\alpha}$ plays the role of a diffusion coefficient for propagation of momentum.

We postulate here that in real (compressible) inhomogeneous fluids this term is caused by hopping of segments from an occupied cage to an adjacent vacant one. Such a postulate was used in the DSCF theory for simple (monomer) fluids of Khan and Shnidman, which produced detailed computations of interfacial and wetting flows. However, the concept of viscous momentum propagation by activated hopping can be traced all the way back to the old kinetic theories of liquids by Eyring [1] and Frenkel [2].

Applying the postulate above to an *inhomogeneous*, *compressible* blend of chains of two types $\alpha = A, B$, interacting with each other and with the walls according to the self-consistent model (37), Eq. (55) is replaced by the following equation for the evolution of momentum density at site $\mathbf{r}$

$$\frac{d\mathbf{g_r}^{\alpha}}{dt} = -\nabla \cdot \left(\mathbf{g_r}^{\alpha}\mathbf{u_r} + \boldsymbol{\varepsilon}_{\mathbf{r}}^{\alpha}\right) - \sum_{k=1}^{12}\frac{\boldsymbol{\pi}_{\mathbf{r},\mathbf{r}+\mathbf{a}_k}^{\alpha}}{a} - \frac{\phi_{\mathbf{r}}^{\alpha}}{wP_{\mathbf{r}}^{\alpha}}\sum_{k=1}^{12}\varsigma_{\mathbf{r},\mathbf{r}+\mathbf{a}_k}^{\alpha}\mathbf{j}_{\mathbf{r},\mathbf{r}+\mathbf{a}_k}^{\alpha} . \tag{57}$$

The terms $\mathbf{g_r}^{\alpha}\mathbf{u_r}$ in Eqs. (55) and (57) represent the convective contributions to the stress, and are identical in form. However, note that now $\mathbf{u_r}$ is the mass-averaged mean velocity, as given by Eq. (14). The term $-\boldsymbol{\varepsilon}_{\mathbf{r}}^{\alpha}$ in Eq. (57) represents the elastic contribution to the stress and is still given by Eq. (53), except that $n_{\mathbf{r}}^{\alpha}$ and $\zeta_{\mathbf{r}}^{\alpha}$ are now functions of $\mathbf{r}$, given by Eqs. (50) and (35), respectively.

The divergence of the viscous stress $\nabla \cdot \left(2\eta_{\mathbf{r}}^{\alpha}\mathbf{D_r}\right)$ in Eq. (55) (which is identical to $\nu_{\mathbf{r}}^{\alpha}\nabla^{2}\mathbf{g_r}^{\alpha}$ according to Eq. (56)) is replaced by the term $\sum_{k}\boldsymbol{\pi}_{\mathbf{r},\mathbf{r}+\mathbf{a}_k}^{\alpha}/a$ in Eq. (57), using a Markov chain model for activated-rate viscous transport of momentum between adjacent sites. The net change $-\boldsymbol{\pi}^{\alpha}_{\mathbf{r},\mathbf{r}+\mathbf{a}_k}/a$ in local momentum density due to hopping of segments of species $\alpha$ across each bond between adjacent sites $\mathbf{r}$ and $\mathbf{r}+\mathbf{a}_k$ is given by

$$
\begin{aligned}
\boldsymbol{\pi}^{\alpha}_{\mathbf{r},\mathbf{r}+\mathbf{a}_k} &= \left(1-\delta_{(\mathbf{r}+\mathbf{a}_k)\cdot\mathbf{e}_3,0}\right)\left(1-\delta_{(\mathbf{r}+\mathbf{a}_k)\cdot\mathbf{e}_3,z_L+\sqrt{2/3}a}\right)\frac{\nu^{\alpha}_{\mathbf{r},\mathbf{r}+\mathbf{a}_k}}{\left(1-\overline{\overline{\phi}}\right)} \\
&\times\left[\mathbf{g}^{\alpha}_{\mathbf{r}}\left(1-\phi^{\alpha}_{\mathbf{r}+\mathbf{a}_k}\right)\varphi\left(\frac{\Delta_{\mathbf{r},\mathbf{r}+\mathbf{a}_k}\left\langle\mathcal{H}^{\alpha}_{\mathbf{r}}\right\rangle}{k_B T}\right)-\mathbf{g}^{\alpha}_{\mathbf{r}+\mathbf{a}_k}\left(1-\phi^{\alpha}_{\mathbf{r}}\right)\varphi\left(-\frac{\Delta_{\mathbf{r},\mathbf{r}+\mathbf{a}_k}\left\langle\mathcal{H}^{\alpha}_{\mathbf{r}}\right\rangle}{k_B T}\right)\right]\frac{\mathbf{a}_k}{a}
\end{aligned}
\tag{58}
$$

For each component of the segmental momentum density vector $\mathbf{g}^{\alpha}_{\mathbf{r}}$, Eq. (58) has the same form as Eq. (29) defining diffusive free segment probability fluxes $\mathbf{j}^{\alpha}_{\mathbf{r},\mathbf{r}+\mathbf{a}_k}$, with components of segmental momentum density $\mathbf{g}^{\alpha}_{\mathbf{r}}$ replacing the free segment probability $P^{\alpha}_{\mathbf{r}}$, and the local kinematic shear viscosity coefficient at zero shear rate $\nu^{\alpha}_{\mathbf{r},\mathbf{r}+\mathbf{a}_k}/\left(1-\overline{\overline{\phi}}\right)$ replacing the local self-diffusion coefficient $D^{\alpha}_{\mathbf{r},\mathbf{r}+\mathbf{a}_k}/\left(1-\overline{\overline{\phi}}\right)$. In a FENE-P dumbbell model for the chain,

$$
\nu^{\alpha}_{\mathbf{r},\mathbf{r}\pm\mathbf{a}_k} = \frac{\zeta^{\alpha}_{\mathbf{r},\mathbf{r}\pm\mathbf{a}_k}\tilde{a}^2_{\alpha}M^{\alpha}_{\mathbf{r}}}{24 m^{\alpha}}
\tag{59}
$$

where $\zeta^{\alpha}_{\mathbf{r},\mathbf{r}\pm\mathbf{a}_k}=\sqrt{\zeta^{\alpha}_{\mathbf{r}}\zeta^{\alpha}_{\mathbf{r}\pm\mathbf{a}_k}}$ is the geometric average of the local segmental friction coefficients at sites $\mathbf{r}$ and $\mathbf{r}+\mathbf{a}_k$ obeying Doolittles' law [69] (Eq. (35)). The transition rate function has been defined in Eq. (31). It depends on $\Delta_{\mathbf{r},\mathbf{r}+\mathbf{a}_k}\left\langle\mathcal{H}^{\alpha}_{\mathbf{r}}\right\rangle/k_B T$, the change the generalized (nonequilibrium) free energy of the system resulting from a segments of type $\alpha$ hopping from site $\mathbf{r}$ to an adjacent vacant site $\mathbf{r}+\mathbf{a}_k$. The dependence of $\left\langle\mathcal{H}^{\alpha}_{\mathbf{r}}\right\rangle/k_B T$ on self-consistent segmental interactions (with segments of opposite type types and with walls), local kinetic energy and chain conformations has been described by Eqs. (36) - (39).

For a one-component, incompressible fluid, $1 - \phi_{\mathbf{r}} = 1 - \overline{\phi}$, and $\varphi(0) = 1$ in Eq. (31), and the kinematic viscosity coefficient $v_{\mathbf{r},\mathbf{r}+\mathbf{a}_k}^{\alpha} = v^{\alpha} = \eta^{\alpha} / \rho^{\alpha}$ is constant. In this case, we recover Newton's law for viscous contribution to the stress $-\boldsymbol{\pi}_{\mathbf{r},\mathbf{r}+\mathbf{a}_k}^{\alpha}$ arising from hops between sites $\mathbf{r}$ and $\mathbf{r} + \mathbf{a}_k$,

$$-\boldsymbol{\pi}_{\mathbf{r},\mathbf{r}+\mathbf{a}_k}^{\alpha} = v^{\alpha} \left( \frac{\mathbf{g}_{\mathbf{r}+\mathbf{a}_k}^{\alpha} - \mathbf{g}_{\mathbf{r}}^{\alpha}}{a} \right) \mathbf{a}_k = \eta^{\alpha} \left( \frac{\mathbf{u}_{\mathbf{r}+\mathbf{a}} - \mathbf{u}_{\mathbf{r}}^{\alpha}}{a} \right) \mathbf{a}_k . \qquad (60)$$

The last term in Eq. (57) has no counterpart in Eq. (55). Its origin can be understood as follows. The probability density for a segment belonging to a chain of connected segments of type $\alpha$ is $\phi_{\mathbf{r}}^{\alpha} / w$. If $\mathbf{j}_{\mathbf{r},\mathbf{r}+\mathbf{a}_k}^{\alpha}$ is non zero, there is a net drift velocity $\mathbf{j}_{\mathbf{r},\mathbf{r}+\mathbf{a}_k}^{\alpha} / P_{\mathbf{r}}^{\alpha}$ due to biased hopping between the sites $\mathbf{r}$ and $\mathbf{r} + \mathbf{a}_k$. This contributes $\left( \phi_{\mathbf{r}}^{\alpha} / w \right) \left( -\zeta_{\mathbf{r},\mathbf{r}+\mathbf{a}_k}^{\alpha} \mathbf{j}_{\mathbf{r},\mathbf{r}+\mathbf{a}_k}^{\alpha} / P_{\mathbf{r}}^{\alpha} \right)$ to the friction force density, and the total friction force density is obtained by summing over all the bonds (it is negligible for an incompressible one-species fluid of ideal chains).

The evolution equation for the total momentum density is obtained by summing over the contributions of the two species,

$$\frac{d\mathbf{g}_{\mathbf{r}}}{dt} = \sum_{\alpha} \frac{d\mathbf{g}_{\mathbf{r}}^{\alpha}}{dt} . \qquad (61)$$

In order to solve the DSCF equations, one has to prescribe boundary conditions for $\mathbf{g}_{\mathbf{r}}$ (or equivalently, for $\mathbf{u}_{\mathbf{r}} = \mathbf{g}_{\mathbf{r}} / \rho_{\mathbf{r}}$) at sites adjoining the walls. We assume that, if a connected segment of type $\alpha$ occupies a site $\mathbf{r}$ adjacent to a wall (which occurs with probability $\phi_{\mathbf{r}}^{\alpha}$), it acquires the velocity of that wall. Therefore, for geometry specified in Figs. 1 and 2, the mass-averaged mean velocity at the sites next to the walls is set to

$$\mathbf{u}_{z=z_1} = -\sum_{\alpha} \phi_{z=z_1}^{\alpha} u_w \mathbf{e}_1, \qquad \mathbf{u}_{z=z_L} = \sum_{\alpha} \phi_{z=z_1}^{\alpha} u_w \mathbf{e}_1, \qquad (62)$$

where $z_1 \equiv \sqrt{2/3}a$, $z_L \equiv \sqrt{2/3}aL$, and L is the number of triangular lattice layers stacked in an FCC lattice arrangement between the walls.

In summary of this section, our formulation of an isothermal DSCF theory for an unentangled blend of two linear homopolymers of types A and B on the FCC lattice results in a system of coupled nonlinear ordinary differential equations for the following set of independent variables defined at each lattice site $\mathbf{r}$: $P_{\mathbf{r}}^A$ and $P_{\mathbf{r}}^B$ (scalar free segment probabilities for each species), $\mathbf{g_r}$ (a 3-component vector of local momentum density), $\mathbf{S}_{\mathbf{r}}^A$ and $\mathbf{S}_{\mathbf{r}}^B$ (second moments of end-to-end distance for ideal chains under flow, second-order symmetric tensors with 6 independent component each). These equations relate first time derivatives of the variables above with nonlinear and nonlocal functions of the same variables. The equations have an identical form for all $\mathbf{r}$, except for sites belonging to boundary layers adjacent to the solid walls parallel to triangular lattice layers stacking the FCC lattice. For sites belonging to these boundary layers, the form of the equations is adjusted to reflect wall impermeability and momentum transfer from moving walls. In the current formulation, periodic boundary conditions are assumed in the directions parallel to the triangular lattice layers. However, they can be easily modified to reflect the presence of other static or moving walls, or some other boundary conditions for exchange of mass, momentum, or both across the boundary.

## III. CONCLUSIONS

We have presented a detailed lattice formulation for a new DSCF theory for inhomogeneous fluids consisting of unentangled homopolymer chains between parallel moving walls. A mean-field factorization of correlations approximates the probability distribution of Kuhn segments by a product of one-body probabilities of free segments in a self-consistent mean field representing interactions with adjacent *connected* segments and walls. The effect of flow on chain conformations is modeled with FENE-P dumbbells, and related to stepping probabilities in a random walk. Free segment and stepping probabilities generate statistical weights for chain conformations in a self-consistent field, and determine local volume fractions of connected segments, which account for intrachain correlations. Flux balance across a unit lattice cell yields deterministic mean-field transport equations for free segment probabilities and momentum densities. Mean-field factorization of correlations suppresses the strong fluctuations associated with a small control volume in a real system or in its MD simulation. Truncated Taylor expansions of the proposed DSCF evolution equations lead to a description by nonlinear, higher order PDEs resembling continuous, deterministic models of transport in interacting polymer fluids derived phenomenologically on the mesoscopic scale.

In an accompanying paper [76], we report a first application of the DSCF theory formulated above to study both transient and steady-states interfacial composition, flows and rheology in unentangled, inhomogeneous polymer fluids. In this study, the fluids are composed of one or two species of linear, flexible homopolymers between solid walls that are sheared at constant, but opposite velocity, and all interfaces are assumed to be planar.

We have presented here a DSCF theory for inhomogeneous fluids consisting of unentangled homopolymer chains. It is based on conservation laws for species occupancies and momentum that are coupled to models of polymer structure and conformation. Since it is formulated on the Kuhn length scale, it is capable of resolving interfacial structure and dynamics at interfaces between phase-separated domains. We note that it can be generalized either by changing the constitutive stress equation (accounting for the reptation regime), the polymer model (e.g., accounting for block copolymers or branching) or by incorporating additional conservation laws and transport equations (e.g., for energy and/or charge). As any mean-field theory, it can also be improved systematically by relaxing the factorization approximation for configurational probabilities, to account for joint probabilities of nearest-neighbor pairs, or of even larger compact clusters. Work along these lines will be reported in future publications.

## ACKNOWLEDGEMENTS

We acknowledge numerous helpful discussions with Drs. Glen H. Ko, Stanislav Solovyov, Wen Tao Li and Dilip Gersappe, for which we are very grateful. We thank Dr. Toshihiro Kawakatsu for sending us a preprint [77] summarizing the relevant work of his group prior to its publication. Most of this work was performed while Y. Shnidman was a member of the Department of Chemical Engineering and Chemistry at the Polytechnic University in Brooklyn, NY, where Maja Mihajlovic and Tak Shing Lo were a Ph. D. candidate and a postdoctoral research associate in his group, respectively. Some preliminary results were included in the Ph. D. thesis [78] submitted by Maja Mihajlovic to Polytechnic University in Brooklyn, NY, in partial fulfillment of requirements for the Ph. D. degree in Chemical Engineering. Maja Mihajlovic and Yitzhak Shnidman thank the National Science Foundation for financial support (grant DMR-0080604). Tak Shing Lo and Yitzhak Shnidman acknowledge a grant from the Mitsubishi Chemical Corporation of Japan, for which they are grateful.

**FIGURES**

**Fig. 1.** Inhomogeneous homopolymer fluids in a channel between two solid walls sheared along the x-axis at opposite velocities. (a) A melt of a single species A. (b) A blend of two species A and B, phase separated into an A-rich phase and a B-rich phase, with planar interfaces parallel to the walls. (c) A sheared drop of a majority-B phase in a majority-A phase matrix.

**Fig. 2.** An FCC lattice site $\mathbf{r} = (x, y, z)$, labeled by $0$, and the distribution of its nearest neighbors among the three triangular lattice planes stacked in the ABC pattern that is characteristic of the FCC lattice.

**Fig. 3.** Random walk and elastic dumbbell representations of chain conformation statistics (for simplicity, a triangular, rather than the FCC lattice, is shown) (a) Chain conformations at equilibrium represented as a lattice random walk (isotropic stepping probabilities denoted by identical bond shading). The isotropic second moment of the end-to-end distance is represented by a sphere. (b) In a velocity field, the stepping probabilities become anisotropic (denoted by different shading of bonds along different lattice directions). The second moment of the end-to-end distance becomes anisotropic as well (represented by an ellipsoid). (c) Interactions of segments with adjacent segments and with walls in an inhomogeneous polymer fluid are represented by random walk in an external self-consistent field (shaded background).

**Fig. 4.** Hopping contributions to the diffusive and viscous center-of-mass fluxes transporting mass and momentum. (a) A segment (denoted by the black solid circle) is shown hopping out of an occupied site enclosed by a control volume (denoted by the rectangular box), into a vacant site (denoted by the white circle). The different shading of circles at the surrounding sites denotes different mean volume fractions that determine the inhomogeneous self-consistent field interacting with the hopping segment. (b) A reverse hop by a segment at the neighboring site into the vacant site enclosed by the same control volume.

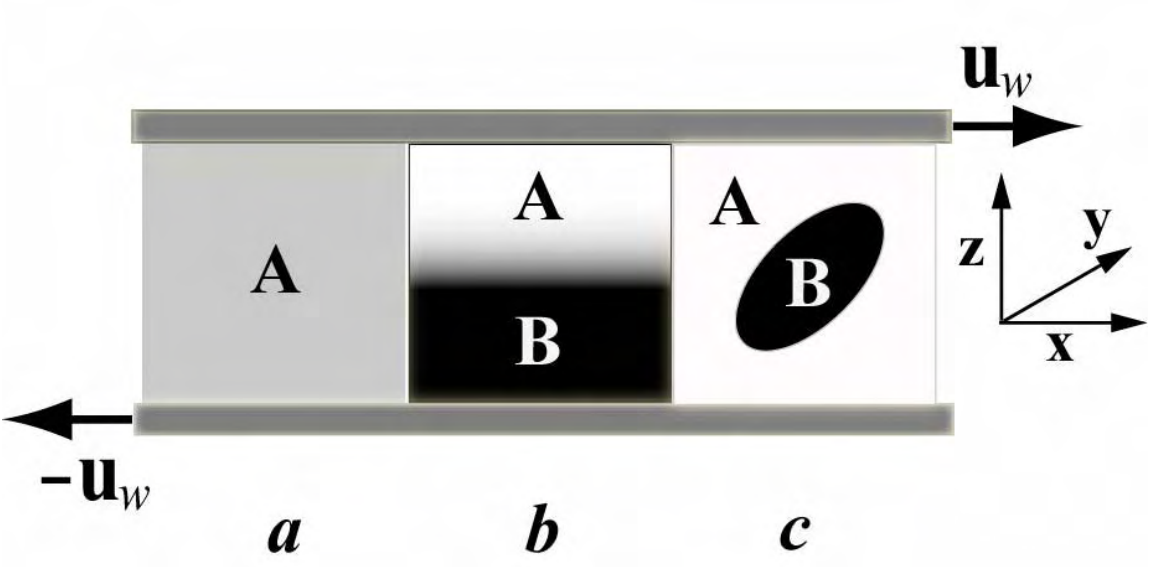

Fig. 2

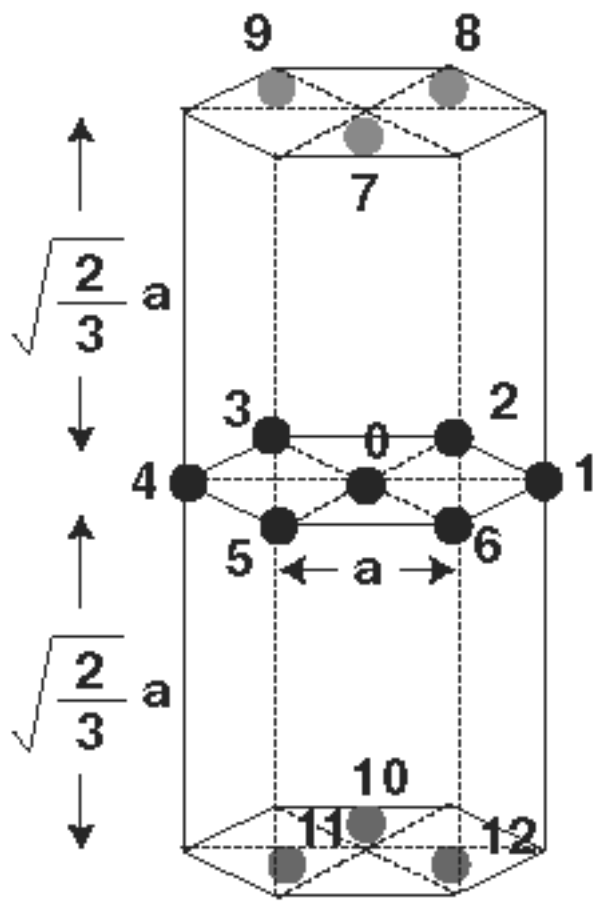

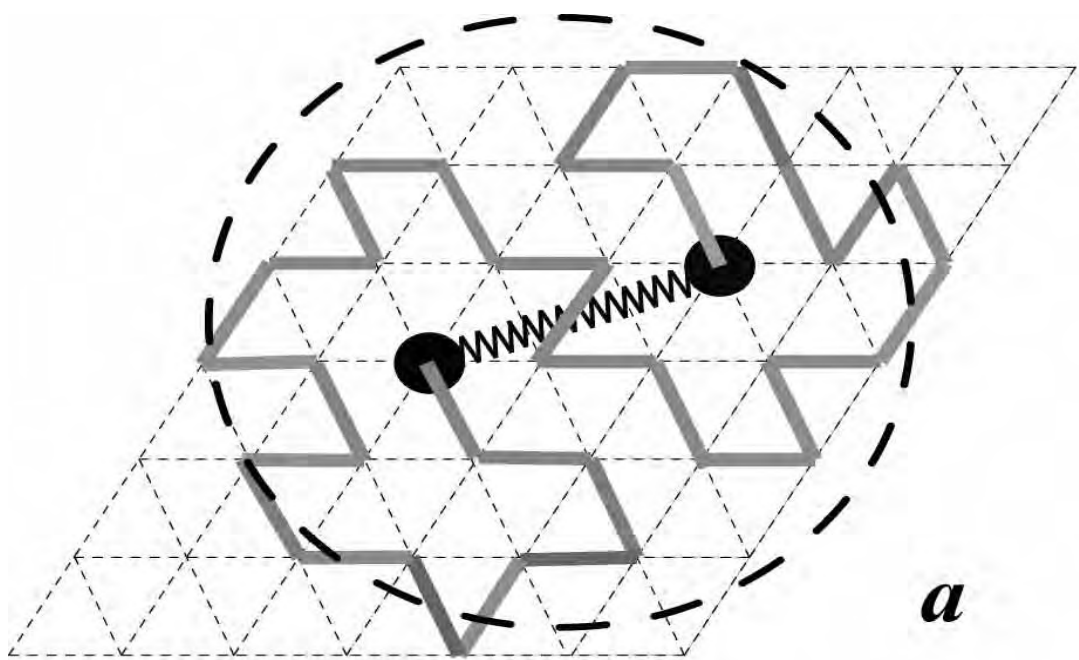

*a*

Fig. 3(a)

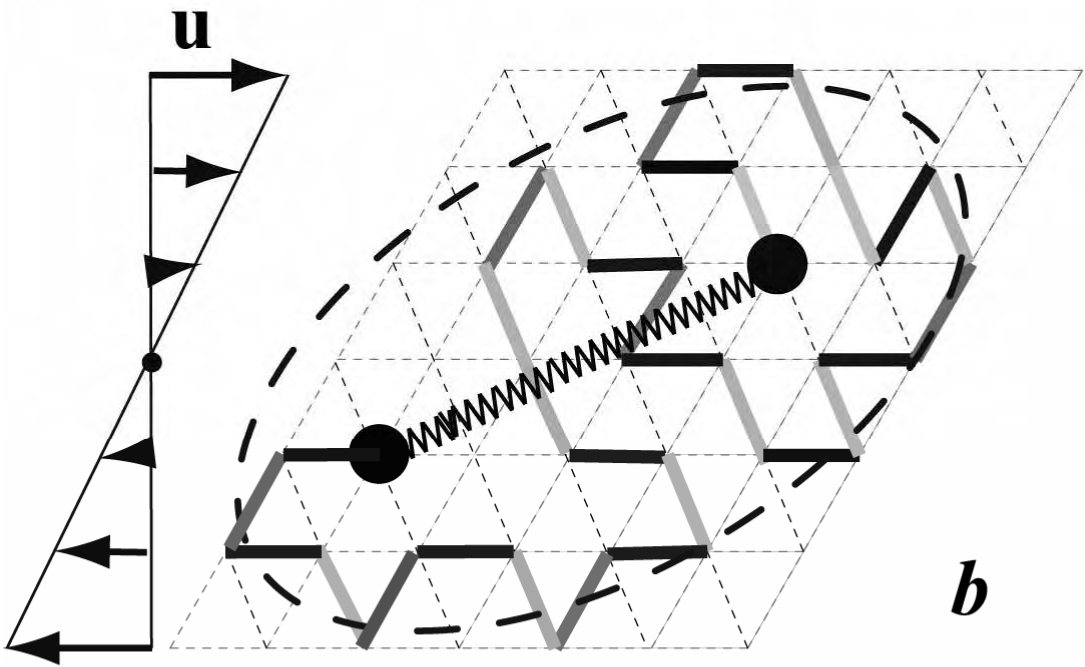

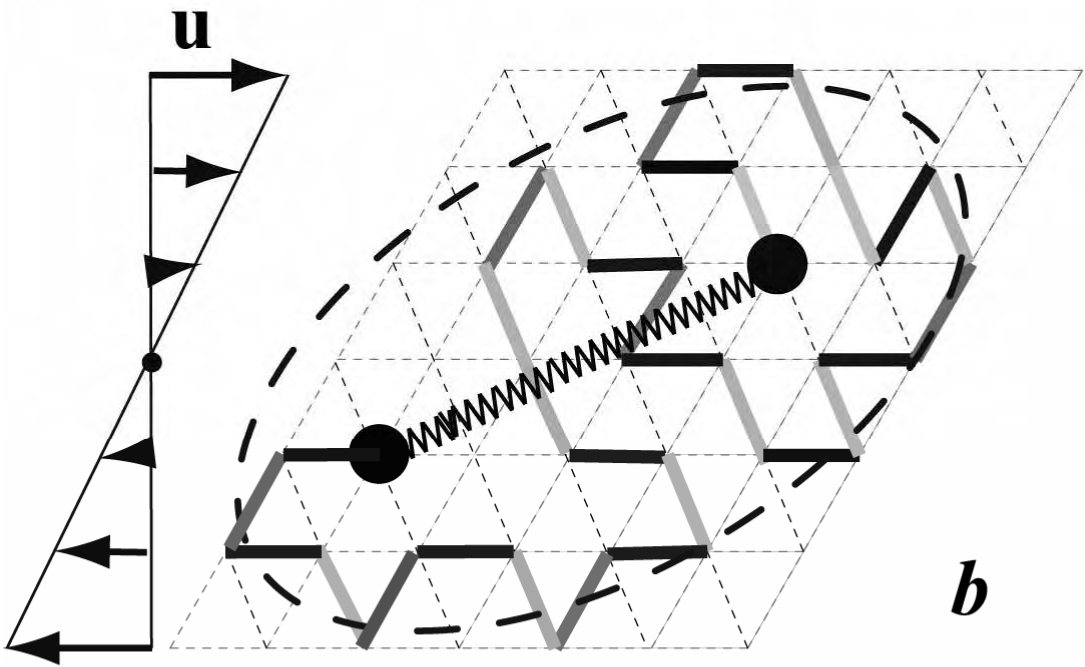

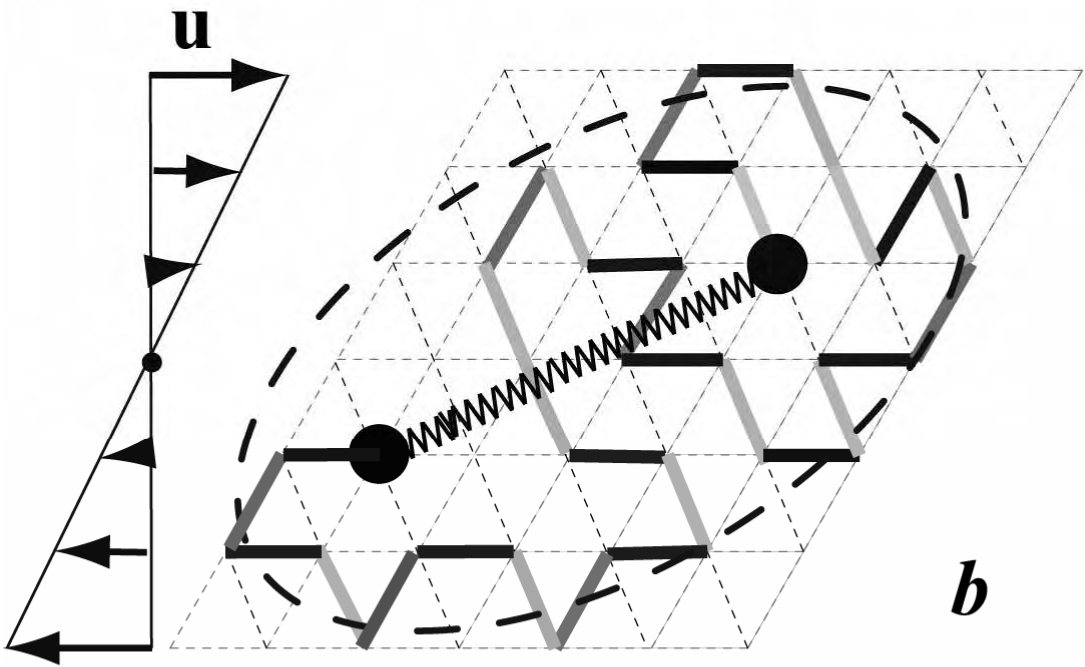

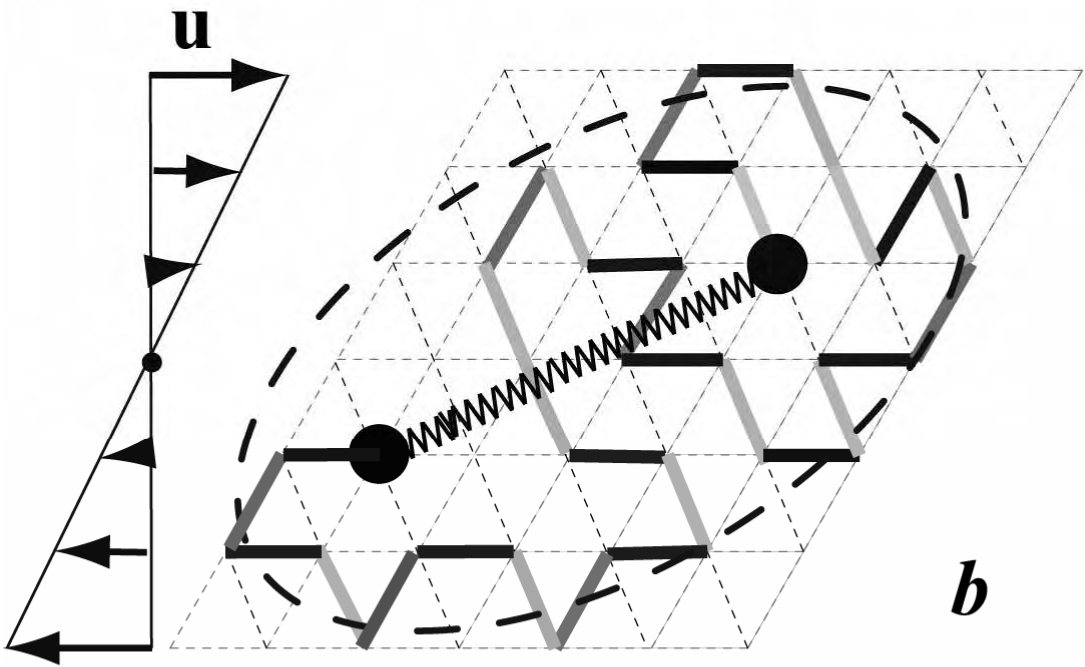

**u**

*b*

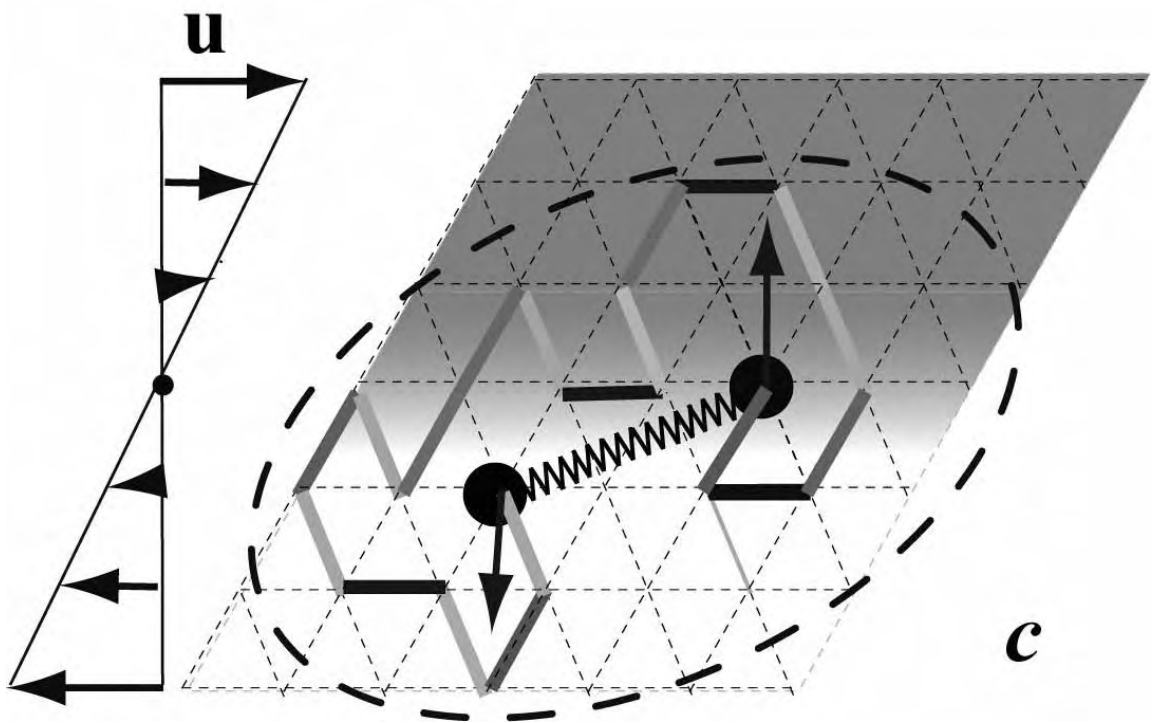

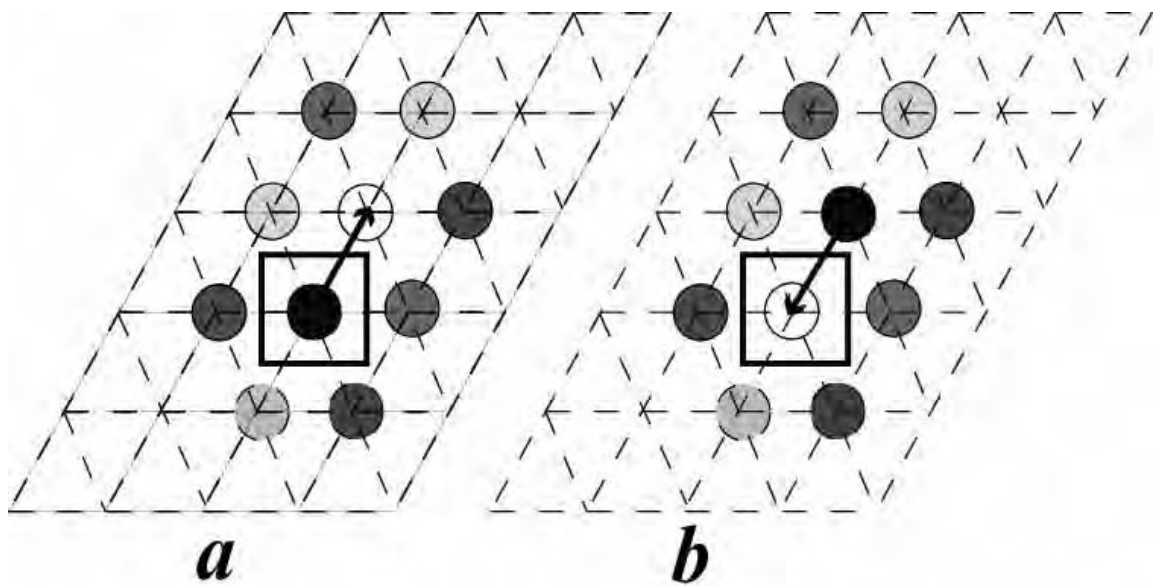

*a*    *b*